\documentclass[reqno,12pt]{amsart}
\usepackage{color} 
\usepackage[pdftex]{hyperref}
\hypersetup{colorlinks=true,linkcolor=red,citecolor=blue,filecolor=magenta,urlcolor=cyan}
\usepackage{times}
\usepackage{amsfonts}
\usepackage{amsmath}
\usepackage{amsthm}
\usepackage{amssymb}
\usepackage{amsbsy}
\usepackage{amscd}
\usepackage{enumerate}
\usepackage{verbatim}
\usepackage{subfigure}

\numberwithin{equation}{section}  
\definecolor{mblack}   {rgb} {0.2, 0.2, 0.2}
\definecolor{mred}     {rgb} {0.6, 0.2, 0.2}
\definecolor{mgreen}   {rgb} {0.2, 0.6, 0.2}
\definecolor{mblue}    {rgb} {0.2, 0.2, 0.6}
\definecolor{mcyan}    {rgb} {0.2, 0.6, 0.6}
\definecolor{mmagenta} {rgb} {0.6, 0.2, 0.6}
\definecolor{myellow}  {rgb} {0.6, 0.6, 0.2}

\newcommand{\leqs}{\leqslant}

\newcommand{\tbar}{{\vert\kern-0.25ex\vert\kern-0.25ex\vert}}
\newcommand{\mvec}{\mathbf}
\newcommand{\mcdomain}{\mathcal{K}}
\usepackage[color=green!20]{todonotes}

\begin{document}

\title[A Note on Optimal Tokamak Control]
      {A Note on Optimal Tokamak Control \\ for Fusion Power Simulation}

\author[M. Holst]{Michael Holst}
\email{mholst@ucsd.edu}
\address{Department of Mathematics\\
         University of California San Diego\\
         La Jolla CA 92093}
\thanks{MH was supported in part by NSF DMS/CM Award 2012857.}

\author[V. Kungurtsev]{Vyacheslav Kungurtsev}
\email{vyacheslav.kungurtsev@fel.cvut.cz}
\address{Department of Computer Science and Engineering\\
         Czech Technical University in Prague\\
         Prague CZ}

\author[S. Mukherjee]{Satwik Mukherjee}
\email{satwik.applied@gmail.com}
\address{Independent Researcher\\
         Hooghly, India}

\date{\today}

\keywords{Magnetohydrodnamics, Navier-Stokes equations, Maxwell equations, Tokamak, Control, Fusion Power Simulation, Petrov-Galerkin Methods, Finite Element Methods}

\maketitle

\begin{abstract}
The Tokamak device is the most promising candidate for producing sustainable electric power by nuclear fusion. 
It is a torus-shaped device that confines plasma by a strong magnetic field.
The development, design and control of the design has been an important area of research, and a significant target is to effectively confine the extremely hot plasma inside its hollow torus-shaped body without touching its boundary for a prolonged period of time.
In an attempt to control a Tokamak device, this paper investigates an optimal control problem for an incompressible, viscous, electrically conducting MHD fluid confined in a closed toroidal region in the presence of an applied current.
The objective functional for the optimal control problem are subject to set of constraint equations, Navier-Stokes and Maxwell equations.
We target the transient control of guiding the plasma to a desired flow at a particular short time instant, where it is expected that the flow had been designed offline to be the desired one from the point of view of steady state operation.
\end{abstract}


\vspace*{-0.2cm}
{\footnotesize
\tableofcontents
}
\vspace*{-0.5cm}

\section{Introduction}

\subsection{Tokamak Modeling and Simulation} \label{intro}
The Tokamak device is the most promising candidate for producing sustainable electric power by nuclear fusion.
It is a torus-shaped device that confines plasma by a strong magnetic field. The development, design and control of the design has been an important area of research, and a significant target is to effectively confine the extremely hot plasma inside its hollow torus-shaped body without touching its boundary for a prolonged period of time. In order to attain a steady state equilibrium flow, magnetic fields are needed in two directions to balance the plasma pressure \cite{Marco}. There are three kinds of coils: Central Solenoid (CS), D-shaped Toroidal Field Coils (TFCs) and Poloidal Field Coils (PFCs) that are used to generate magnetic fields inside the Tokamak. The alternating current density through central solenoid coil (which works as a primary transformer) along with several D-shaped coils create
an alternating magnetic field in plasma inside the Tokamak, thus producing a high electric current in the plasma. To maintain the stability of the plasma, robust feedback control is necessary to manipulate the magnetic field to target a desired steady state position, shape and current of the plasma.
This applied current density parameter is one of such important feedback control parameter which can regulate magnetic field, shape, various uncertainties and disturbances arising in plasma flow. 

In an attempt to control a Tokamak device, this paper investigates an optimal control problem for an incompressible, viscous, electrically conducting MHD fluid confined in a closed toroidal region in the presence of an applied current. The objective functional for the optimal control problem are subject to set of constraint equations, Navier-Stokes \cite{GUNZBURGER2007295} and Maxwell equations \cite{YOUSEPT}. We target the transient control of guiding the plasma to a desired flow at a particular short time instant, where it is expected that the flow had been designed offline to be the desired one from the point of view of steady state operation. 

In addition, we will study the effect of physical uncertainty in the control model in the form of stochastic partial differential equation (PDE) systems with parameters defined to be noisy and unknown. In this paper, electrical resistivity in plasma is considered as a stochastic uncertain parameter. In most of the previous work, the electrical conductivity parameter in plasma was assumed a fixed parameter; however, according to Marco and Pironti \cite{Marco}, the assumption of known electrical conductivity $\sigma_{p}$ in plasma is not rigorously correct, thereby demanding to its more general treatment in a Tokamak model. In this context, we would like to add that, an optimization problem with such stochastic PDE constraints are also related to the inverse problems as addressed by several researchers in the literature \cite{article1,ZABARAS20084697}. Additionally, we have also considered kinematic viscosity of plasma as an uncertain parameter, as described in the work of \cite{alfven1}. For a methodical pedagogy, the use of stochastic finite element methods to solve such stochastic PDE systems have been found to be very useful, and research papers with such a methodology are extensively available in the literature \cite{Xiu}; however, the study of stochastic control problems constrained by stochastic PDE systems are less common \cite{HOU201187,Borz,ROSSEEL2012152}, hence its application in magnetic Tokamak control problem is a novelty of this work.  

\subsection{Related Work}
The Tokamak device has been modeled by exploiting its axis symmetry along the toroidal angular direction, as can be found in \cite{alfven1, alfven2, alfven3}. This works introduced a new set of coordinate axis-symmetric transformations that turned out to be very useful to describe the cross-sectional part of a Tokamak. However, their study was limited to analyze the steady state behavior of plasma inside a Tokamak. A similar analytical approach was also mentioned in the book \cite{Marco}. Apart from the analytical studies, experimentally motivated papers on Tokamak and Stellarator that are very important for Modelling a Tokamak are well existed in the literature \cite{AGREDANOTORRES2021112683,Mukhovatov_1971,Poli, ZhaoLin, Sizyuk,Rozhansky_2001, Stacey,en3111741, GTC}. The existence of turbulence on the outer part of Tokamak plasma brings hindrance against achieving the steady stable plasma state, thereby attracting a significant number of researchers to address this issue in their relevant papers \cite{GTC, Sizyuk,Mukhovatov_1971}. As the Tokamak plasma is kept at an extremely high temperature, several uncertainty regarding intrinsic plasma properties such as its thermal conductivity, electrical conductivity starts to change in an unpredictable manner, thereby generating scope for analyzing such a phenomena borrowing the tools and techniques from uncertainty quantification and reinforcement learning as can be found in \cite{Degrave,Coster,mhd,Sullivan}. The techniques of uncertainty quantification is prevalent in exploring the underlying irregularities and hidden nature of its intrinsic physical dynamical features, thereby employing such tools in optimally control a very complex engineering device such as a Tokamak is also a very engaging field of research as can be found in the literature \cite{10.1007/978-3-319-55795-3_1,RASTOVIC2007676, Tiesler,kessel,GARRIDO2009431,Lakhlili,tokamaks}.

Suitable normed linear spaces are formulated to accommodate the solutions of the optimal control minimizing problem. We use a non-recursive one-shot approach to the formulated control problem and utilize the stochastic collocation method to solve the resulting set of equations. The remaining part of the paper is organized in the following manner. Section \ref{notations} describes and defines the required normed linear space where the weak solutions of the control problem lies. The mathematical modeling of a Tokamak is carried out in Section \ref{problem}. Moreover, we demonstrate the existence of weak solutions for the chosen set of equations describing the Tokamak model, applying classical results in the literature. In Section \ref{stochastic}, we delve into the formulation of the stochastic version of the problem by introducing the electrical conductivity of plasma as a stochastic parameter. 
\subsection{Mathematical Notation and Definitions} \label{notations}
We will use standard mathematical notation for the Navier-Stokes, Maxwell, and MHD equations, as well as for the relevant spaces of functions and vector fields involved, as used for example in~\cite{OriginalEq,HLT08a,HoTi96b,BeHo18b}.

To begin, for any coordinate basis such as $(a,b,c)$ of $\mathbb{R}^3$ we denote the unit basis vectors by $\hat{i}_a,\hat{i}_b,\hat{i}_c$, respectively. 
A scalar time-dependent function $u \colon \mathbb{R} \to \mathbb{R}^3$ is denoted using normal font as $u(t)$, with bold font $\mvec{u}(t)$ indicating a generically time dependent vector field $\mvec{u}(t)=u_a(t)\hat{i}_a+u_b(t)\hat{i}_b+u_c(t)\hat{i}_c$.
The Poisson bracket will be denoted as $\{ u, v \} = \frac{\partial u}{\partial a} \frac{\partial v}{\partial b} - \frac{\partial u}{\partial b} \frac{\partial v}{\partial a} $.
 Let $\Omega \subset \mathbb{R}^n$ be an open subset of $\mathbb{R}^n$.
The Lebesgue integrable spaces of scalar functions over $\Omega$ are defined as
\begin{align*}
    L^p(\Omega) & = \left\{~ u \colon \Omega \to \mathbb{R} ~\big|~ \|u\|_{p, \Omega} < \infty ~\right\}, \end{align*}
    where the $L^p$ norms are defined as:
    \begin{align*}
    \| u \|_{p,\Omega} &= \left(\int_{\Omega} |u|^p ~dx \right)^{1/p}, \quad 1 \leqs p < \infty,
\qquad
    \| u \|_{\infty,\Omega} = \sup_{x \in \Omega} |u|
    \end{align*}
The case $p=2$ allows also for an inner-product:
    \begin{equation}
(\phi,\psi)_2 =
   (\phi,\psi)_{L^{2}(\Omega)} 
   = \int_{\Omega} \phi \psi ~dx. 
\end{equation}
The Sobolev spaces $W^{k,p} (\Omega)$ are a family of Banach spaces characterized by
\begin{align}
W^{k,p} (\Omega) & = \left\{~ u \in L^2(\Omega) ~:~ \|u\|_{k,p,\Omega} < \infty ~\right\},
\end{align} 
where the Sobolev norms are defined as
\begin{align}
\|u\|_{k,p,\Omega} 
   & = \|u\|_{W^{k,p}(\Omega)}
     = \Big( \sum_{0 \leqs |\alpha|\leqs k} \|D^{\alpha}u\|_p^p \Big)^{1/p},
\end{align} 
with
\begin{align}
D^{\alpha}u = \frac{\partial^{|\alpha|} u}
                   { \partial x_1^{\alpha_1}
                     \partial x_2^{\alpha_2}
                     \cdots
                     \partial x_n^{\alpha_n} },
\qquad
|\alpha| = \sum_{j = 1}^{n} \alpha_j,
\end{align}
where $\alpha=(\alpha_1,\alpha_2,\ldots,\alpha_n)$ is a multi-index with each $\alpha_k \in \mathbb{W}$ (non-negative integers).
Note that the case of $k=0$ reduces to $L^p(\Omega)=W^{0,p}(\Omega)$.
The case of $p=2$ are a family of Hilbert spaces with appropriately defined inner-products that induce the norms above. 
The singular case of $p=2$ and $k=1$ is particularly important for second order PDE systems, and is often denoted more simply as
$H^1(\Omega) = W^{1,2}(\Omega)$.
The important subspace of $W^{1,p}(\Omega)$ that vanishes on the boundary is denoted
\begin{align}
W^{1,p}_0(\Omega) = \{~ u \in W^{1,p}(\Omega) ~:~ u = 0 \mbox{ on } \partial \Omega \mbox{ (in trace sense) } ~\},
\end{align}
with again the special case of $p=2$ denoted
$H^1_0(\Omega) = W^{1,2}_0(\Omega)$.
Extensions of these normed and inner-product spaces of scalar functions to vector (and more generally, tensor) fields come naturally through composition with the discrete $p$-norms of finite-dimensional vectors.
A common notation is the use of bold font to denote the spaces of vector and tensor fields, e.g., $\mathbf{L}^p(\Omega)$ and $\mathbf{W}^{k,p}(\Omega)$.
Note that $f = (f_1,f_2,\cdots,f_k) \in \mathbf{L}^p(\Omega)$ if and only if each of the component functions satisfy $f_{i} \in L^p(\Omega)$. 

The set of all infinitely differentiable scalar, vector, and tensor functions with compact support in $\Omega$ is denoted by $\mathcal{D}(\Omega)$.
The divergence-free smooth functions are denoted as,
\begin{eqnarray} \label{tokdefn2}
    \mathcal{\tau^{\infty}}(\Omega) & = & \left\{~ v \in \mathcal{D}{(\Omega)} 
    ~:~ \nabla \cdot v = 0 ~\right\}, \end{eqnarray}
The following sub-spaces are introduced for weakly divergence-free functions in the distributional sense,
    \begin{eqnarray} \label{tokdefn22}
    \mathcal{\tau}(\Omega) & = & \left\{~ v \in L^{2}{(\Omega)} ~:~ \nabla \cdot v =  0 ~\right\},
\end{eqnarray}
where $\nabla \cdot v = 0$ in the weak sense, meaning that $\int_{\Omega} v \cdot \nabla \phi = 0, \forall \phi \in \mathcal{D} (\Omega)$. 
The closure of $\tau^{\infty} (\Omega)$ in $L^2(\Omega)$ is denoted by $\overline{\tau}(\Omega)$. 
We have $\overline{\tau}(\Omega) \subset \tau(\Omega) \subset L^2(\Omega)$. 
Similarly, for the Sobolev spaces, we define, $\tau_{p}^{k}(\Omega) = W^{k,p}(\Omega) \bigcap \tau (\Omega) $ and $\overline{\tau_{p}^{1}}$ denotes the closure of $ \tau^{\infty} (\Omega) $ in the $W^{1,p}(\Omega)$. 
As we will see below, solutions of the optimization problem will be sought in the spaces
\begin{eqnarray}
\widetilde{\tau_{2}^{2}} (\Omega) & = & \left\{~ v \in \tau_{2}^{2} (\Omega) ~:~ (v \cdot n)|_{\partial \Omega} = 0, ~ (\nabla \times v)|_{T_S}|_{\partial \Omega} = 0 ~\right\},
\end{eqnarray}
where $T_S$ denotes the tangent plane at Tokamak surface (see below).
The closure of $\widetilde{\tau_{2}^{2}} (\Omega)$ inside $W^{1,2}(\Omega)$ is denoted by $\overline{\tau_{2}^{1}}$.

\begin{figure}
    \centering
    \includegraphics[height=3.0in]{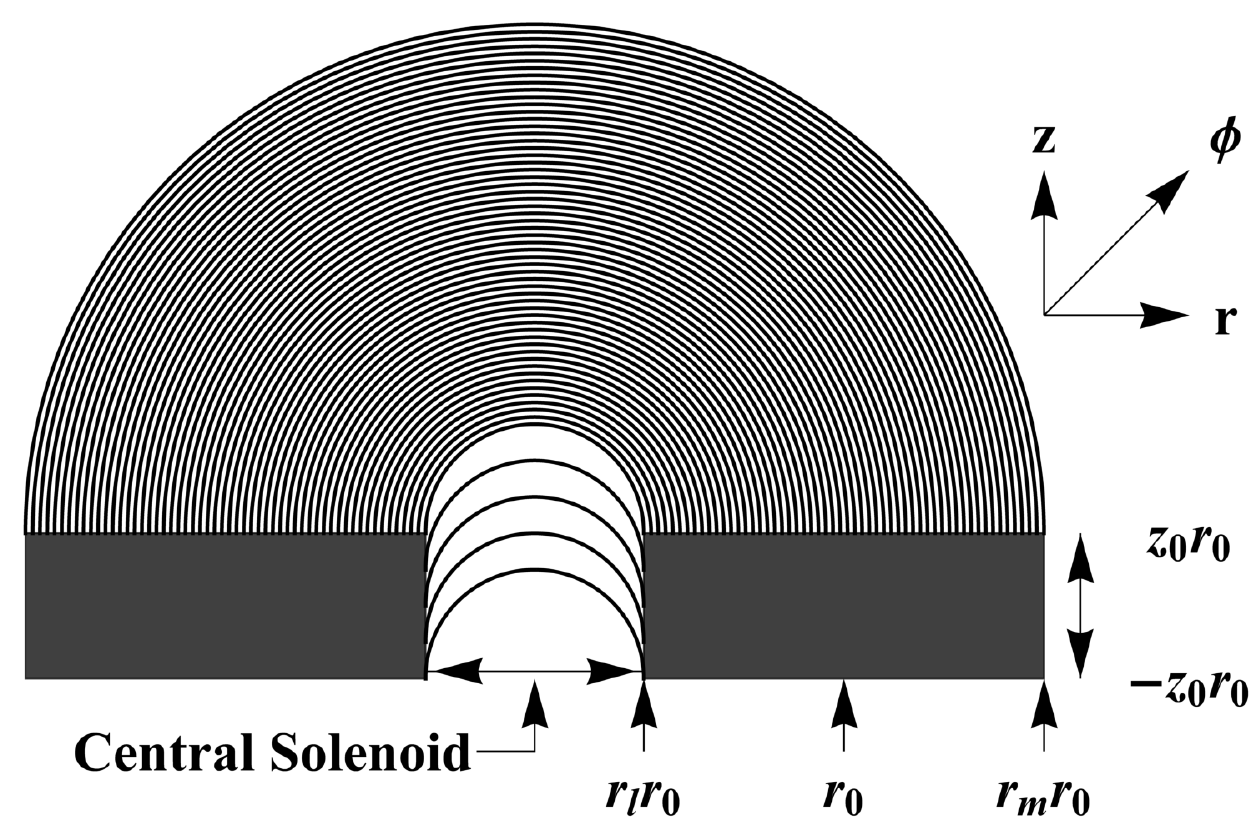}
    \caption{\small Physical sketch of an inner cross-section of TOKAMAK. Here, $-z_{0} r_{0},z_{0} r_{0}$ denote its height. $r_{l} r_{0}$ denotes its left most part and $r_{m} r_{0}$ denotes its right most part. The empty middle space contains a central solenoid magnet which is responsible for Ohmic heat generation in plasma.}
    \label{Diagram}
\end{figure}

\section{System Formulation} \label{problem}
The mathematical description of the optimal control problem for a Tokamak is presented in this section. The Tokamak is denoted by a torus shaped bounded region $\Omega \subset \mathbb{R}^3$, which contains plasma. The boundary wall of Tokamak is denoted by $\partial \Omega$, which is the outer surface of a torus. A physical sketch of Tokamak is presented in Figure \ref{Diagram}. The plasma is modeled as a magnetohydrodynamic (MHD) fluid \cite{alfven1, alfven2, alfven3}. Let $\mvec{v}$ denotes the velocity field of plasma and $\mvec{B}$ the induced magnetic field. The magnetic field $\mvec{B}$ is induced due to the high current density in external Tokamak coils. The current density in plasma is denoted by $\mvec{J_p}$. Plasma experiences a strong Lorentz force due to migration of electric charges in the presence of a strong magnetic field, which acts on the current as $\mvec{E} + \mvec{v} \times \mvec{B}$ where $\mvec{E}$ is the electric field in plasma. We now present the ideal MHD equations governing the plasma flow inside the Tokamak.

\subsection{Classical Formulation}

To begin, we divide the electric $\mvec{E}$ and magnetic $\mvec{B}$ fields inside the plasma into endogenous and induced components:
\begin{align*}
\mvec{E} &= \mvec{E}_0 + \mvec{E}_{\kappa (t)}^{ind} \\
\mvec{B} &= \mvec{B}_0 + \mvec{B}_{\kappa (t)}^{ind} 
\end{align*}
The induced component is taken to be controlled with precision by the engineer operating the current in the coils surrounding the Tokamak; we denote as $\mvec{E}_{\kappa (t)}^{ind}$ the externally controlled induced electric field and $\mvec{B}_{\kappa (t)}^{ind}$ the induced magnetic field, with $\mvec{E}_0$ and $\mvec{B}_0$ the endogenous electric and magnetic fields, respectively.

We assume that the dynamical time scale of the energy and current diffusion is much smaller than the duration of a discharge, so it is reasonable to discard the time derivative of the electric field $\mvec{E_{t}}$ in Maxwell's equations as trivially small. We write the electrohydromagnetic field laws to describe plasma inside the Tokamak on $\Omega \times [0,T]$ as follows:
\begin{flalign}
\label{eq:max2}
   \mvec{v}_{t} 
     + (\mvec{v} \cdot \nabla) \mvec{v} 
     - \nu {\nabla}^{2} \mvec{v}
     - \mvec{J}_{p} \times \mvec{B} 
     + \nabla p
   & = 0,
   & \textrm{(\textit{Force Balance})} \\
\label{eq:div}
   \nabla \cdot \mvec{v} 
   & =  0 , 
   & \textrm{(\textit{Incompressibility})} \\
\label{eq:max4}
   \nabla \cdot \mvec{B}
   & =  0, 
   & \textrm{(\textit{Transversality})} \\
\label{eq:max3}
   \nabla \times \mvec{B} 
   & = \mvec{J}_{p}, 
   & \textrm{(\textit{Ampere's Law})} \\
\label{eq:m1}
   \nabla \times \mvec{E} 
     + \mvec{B}_t
   & = 0,
   & \textrm{(\textit{Faraday's Law})} \\
\label{eq:lorentz}
   \mvec{E} 
     + \mvec{v} \times \mvec{B} 
   & =  \eta \mvec{J_{p}},
   & \textrm{(\textit{Ohm's Law})}
\end{flalign}
for some finite time $T>0$, 
where $\mvec{J_{p}}$ denotes the current density,
$p$ denotes plasma pressure,
$\nu$ denotes kinematic viscosity, and
$\eta$ denotes the electrical resistivity. 
In more general formulations of MHD, there are two additional physical 
parameters: the magnetic permeability (typically denoted $\mu$) and
the electronic conductivity (typically denoted $\sigma$). 
In this work, as is done in much of the Tokamak literature, 
we simplify the presentation by considering them to be both one. 

The set of initial and boundary conditions completing the system is given below:
\begin{align} \label{eq:rectbound}
\mvec{v}(0) &= \mvec{v}_1, & \text{on } \Omega  \nonumber\\
\mvec{B}(0) &= \mvec{B}_1, & \text{on } \Omega  \nonumber \\
\mvec{v}(t) &= 0, & \text{on }\partial \Omega \times [0,T] \\
\mvec{B}(t) \cdot \mvec{n} &= 0, & \text{on }\partial \Omega \times [0,T]\nonumber\\
 \nabla \times \mvec{B}  &= 0, & \text{on }\partial \Omega \times [0,T] \nonumber
\end{align}
indicating initial conditions and, for safe operation, no active fields transverse to the Tokamak boundary.

\subsection{Weak Formulation}

We note that Ampere's Law and Ohm's Law together imply
\begin{align}
\eta \nabla \times \mvec{B} = \mvec{E} + \mvec{v} \times \mvec{B}.
\end{align}
Taking the curl of both sides gives
\begin{align}
\eta \nabla \times \nabla \times \mvec{B} 
   = \nabla \times \mvec{E} + \nabla \times (\mvec{v} \times \mvec{B}).
\end{align}
Using the simple product formula
\begin{align}
\nabla \times (\mvec{v} \times \mvec{B}) 
   = \mvec{B} \cdot \nabla \mvec{v} - \mvec{v} \cdot \nabla \mvec{B},
\end{align}
we then have
\begin{align}
\eta \nabla \times \nabla \times \mvec{B} 
   = \nabla \times \mvec{E} 
   + \mvec{B} \cdot \nabla \mvec{v} - \mvec{v} \cdot \nabla \mvec{B}.
\end{align}
Using now Faraday's law we have produced the reduced Maxwell equation
\begin{align}
\mvec{B}_t
   + \eta \nabla \times \nabla \times \mvec{B} 
   + \mvec{v} \cdot \nabla \mvec{B}
   - \mvec{B} \cdot \nabla \mvec{v} 
   = 0.
\end{align}
This leads to the following reformulation of 
\eqref{eq:max2}--\eqref{eq:lorentz} as
the MHD system:
Find $\mvec{v} \in \mathcal{V}$ 
and $\mvec{B} \in \mathcal{B}$, such that
\begin{flalign}
\label{eq:mhd-eq1}
   \mvec{v}_{t} 
     + (\mvec{v} \cdot \nabla) \mvec{v} 
     - \nu {\nabla}^{2} \mvec{v}
     + \mvec{B} \times (\nabla \times \mvec{B})
     + \nabla p
   & = 0,
   \\
\label{eq:mhd-eq2}
   \nabla \cdot \mvec{v} 
   & = 0, 
   \\
\label{eq:mhd-eq3}
   \mvec{B}_t
   + \eta \nabla \times (\nabla \times \mvec{B})
   + \mvec{v} \cdot \nabla \mvec{B}
   - \mvec{B} \cdot \nabla \mvec{v} 
   & = 0,
   \\
\label{eq:mhd-eq4}
   \nabla \cdot \mvec{B} 
   & = 0,
\end{flalign}
subject to the initial and boundary conditions~\eqref{eq:rectbound},
where the solution spaces $\mathcal{V}$ and $\mathcal{B}$
containing $\mvec{v}$ and $\mvec{B}$ are the (possibly different)
divergence-free subspaces of the suitably chose
time-parameterized (Bochner) Sobolev spaces:
\begin{flalign}
\mathcal{V} &= \left\{~ \mvec{v} \colon \Omega \times [0,T] \to \mathbb{R}^3 
                     ~|~ \nabla \cdot \mvec{v} = 0 ~\right\}
   \subset L^r([0,T],W^{k,p}(\Omega)),
\\
\mathcal{B} &= \left\{~ \mvec{B} \colon \Omega \times [0,T] \to \mathbb{R}^3 
                     ~|~ \nabla \cdot \mvec{B} = 0 ~\right\}
   \subset L^s([0,T],W^{m,q}(\Omega)).
\end{flalign}

A weak formulation of~\eqref{eq:mhd-eq1} arises through requiring orthogonality of the residual with respect to a test function $\mvec{w} \in \mathcal{W}$ in the sense of the extended $L^2$-inner-product, for suitable test space $\mathcal{W}$.
I.e., we seek $\mvec{v} \in \mathcal{V}$, such that
\begin{flalign}
(\mvec{v}_{t} + (\mvec{v} \cdot \nabla) \mvec{v},\mvec{w})_2
     - \nu ({\nabla}^{2} \mvec{v}, \mvec{w})_2
     + (\mvec{B} \times (\nabla \times \mvec{B}),\mvec{w})_2
     + (\nabla p,\mvec{w})_2
   & = 0,
\label{eq:mhd-weak-eq1}
\end{flalign}
$\forall \mvec{w} \in \mathcal{W}$.
A weak formulation of~\eqref{eq:mhd-eq3} arises similarly, for a test function $\mvec{z} \in \mathcal{Z}$ for suitable test space $\mathcal{Z}$.
I.e., we seek $\mvec{B} \in \mathcal{B}$, such that
\begin{flalign}
   (\mvec{B}_t,\mvec{z})_2
   + (\eta \nabla \times (\nabla \times \mvec{B}),\mvec{z})_2
   + (\mvec{v} \cdot \nabla \mvec{B},\mvec{z})_2
   - (\mvec{B} \cdot \nabla \mvec{v} ,\mvec{z})_2
   & = 0,
\label{eq:mhd-weak-eq2}
\end{flalign}
$\forall \mvec{z} \in \mathcal{Z}$.
The function spaces $\mathcal{W} \subset L^2(\Omega)$ and $\mathcal{Z} \subset L^2(\Omega)$ are Banach spaces that must be chosen so that each term in~\eqref{eq:mhd-weak-eq1}--\eqref{eq:mhd-weak-eq2} is finite for all possible arguments from $\mathcal{W}$ and $\mathcal{Z}$.
Sufficient conditions on $\mathcal{W}$ and $\mathcal{Z}$ that will guarantee this behavior are easily determined through use of H\"{o}lder inequalities in $L^p$ spaces and other standard norm inequalities in Sobolev spaces (cf.~\cite{BeHo18b,BeHo15a}).
This choice for $\mathcal{W}$ and $\mathcal{Z}$ then produces 
a well-defined weak formulation:
Find $(\mvec{v},\mvec{B}) \in \mathcal{V} \times \mathcal{B}$ such that
\eqref{eq:mhd-weak-eq1}--\eqref{eq:mhd-weak-eq2} hold for all
     $(\mvec{w},\mvec{z}) \in \mathcal{W} \times \mathcal{Z}$.
After a well-defined weak formulation is produced through the careful
choice of the four Banach spaces 
$\mathcal{V},\mathcal{B},\mathcal{W},\mathcal{Z}$,
it remains to prove that there actually exists a weak solution 
$(\mvec{v},\mvec{B}) \in \mathcal{V} \times \mathcal{B}$,
and hopefully that this solution is unique in $\mathcal{V} \times \mathcal{B}$.
In other words, one hopes that the four spaces have been chosen together so that the solution spaces $\mathcal{V}$ and $\mathcal{B}$ are large enough to contain a (weak) solution $(\mvec{v},\mvec{B})$, but not so large that there is more than one solution.

\subsection{A Priori Estimates and Well-Posedness}

The system~\eqref{eq:mhd-eq1}--\eqref{eq:mhd-eq4} is a special case of a more general MHD system studied in~\cite{OriginalEq} (see also~\cite{GUNZBURGER2007295,HLT08a,HoTi96b}).
The analysis in~\cite{OriginalEq} is centered on a generalization of the weak formulation~\eqref{eq:mhd-weak-eq1}--\eqref{eq:mhd-weak-eq2} above, allowing for general magnetic permeability and electric conductivity, and working with variables $\mvec{E}$ (the electric field) and $\mvec{H}$ (the magnetizing field) rather than $\mvec{E}$ and $\mvec{B}$ (the magnetic field), with $\mvec{H}$ and $\mvec{B}$ related through the (possibly non-constant) magnetic permeability:
$$
\mvec{B} = \mu \mvec{H}.
$$
The analysis in~\cite{OriginalEq} proceeds by first establishing some standard energy estimates for the $L^2$-norms of the plasma velocity $\mvec{v}$ and the magnetizing field $\mvec{H}$, based on a combination of norm estimates and use of a Grownwall lemma for differential inequalities.
Subsequently, $L^2$-estimates for $\mvec{v}_t$, of some spatial derivatives of $\mvec{v}$, and of the pressure gradient $\nabla p$ are obtained using similar arguments.
This then leads to a uniqueness result (Theorem 3.1 in~\cite{OriginalEq}) for a generalization of~\eqref{eq:mhd-eq1}--\eqref{eq:mhd-eq4}, under initial and boundary conditions that include~\eqref{eq:rectbound}.

Finally, an existence result (Theorem 4.1 in~\cite{OriginalEq}) for a generalization of~\eqref{eq:mhd-eq1}--\eqref{eq:mhd-eq4} are established in~\cite{OriginalEq}, under initial and boundary conditions that include~\eqref{eq:rectbound}.
The proof is based on the use of the Galerkin method, whereby a sequence of Galerkin approximations to solutions of~\eqref{eq:mhd-weak-eq1}--\eqref{eq:mhd-weak-eq2} (or rather, to a more general version of our system) is generated, and the \emph{a priori} estimates established earlier are shown to hold for each Galerkin approximation.
The sequence is then shown to converge in the limit, and this limit is subsequently shown to satisfy (the generalization of)~\eqref{eq:mhd-weak-eq1}--\eqref{eq:mhd-weak-eq2}.
Therefore, thanks to the analysis in~\cite{OriginalEq} of a generalization 
of our weak formulation~\eqref{eq:mhd-weak-eq1}--\eqref{eq:mhd-weak-eq2},
there exists a unique weak solution 
to~\eqref{eq:mhd-weak-eq1}--\eqref{eq:mhd-weak-eq2},
and under suitable regularity assumptions, to the strong formulation
\eqref{eq:mhd-eq1}--\eqref{eq:mhd-eq4}
as well as to the original classical formulation
\eqref{eq:max2}--\eqref{eq:lorentz}.

\section{Optimal Control Problem}
\subsection{Cylindrical Coordinate Formulation}

The complex structure of Tokamak is often modeled in the axis symmetric cylindrical coordinates system $(r,\phi,z)$, where $r$ denotes the radial direction, $z$ the Tokamak height and $\phi$ the toroidal angle direction. With respect to the standard coordinate basis, we can write a substitution as by,
\[
x=r\cos \phi, \quad y=r\sin\phi, \quad z=z.
\]
The velocity $\mvec{v}=(v_{r}, v_{\phi}, v_{z})$ and magnetic fields $\mvec{B}  = (B_{r}, B_{\phi}, B_{z})$ inside the Tokamak are assumed to be symmetric with respect to the angle $\phi$, i.e. $\frac{\partial \mvec{B} + \mvec{B}_{\kappa (t)}^{ind}}{\partial \phi} =\frac{\partial \mvec{v}}{\partial \phi} = \mvec{0}$. It is standard to exploit this symmetry appearing in the form and dynamics of the Tokamak by performing a coordinate transformation of the system of equations above into cylindrical coordinates. 

To present the transformation we first define two flux quantities as follows:
\begin{align} \label{eq:chi}
    \psi(r,z) = \frac{1}{2 \pi} \int_{S(r)} \mvec{v} \cdot dS, 
    \qquad
    \bar{\chi}(r,z) &= \frac{1}{2 \pi} \int_{S(r)} \left(\mvec{B} + \mvec{B}_{\kappa (t)}^{ind} \right) \cdot dS, \nonumber \\
    \bar{\chi} = \chi + \kappa (t), ~ \kappa \colon \mathbb{R} \to \mathbb{R} 
    \qquad
    \chi(r,z) &= \frac{1}{2 \pi} \int_{S(r)} \mvec{B} \cdot dS.
\end{align}
Here, $S(r)$ denotes the surface of a sphere with radius $r$, with the center at the origin in the torus interior of the Tokamak, and radius typically confined to correspond to the inside of the toroidal device. These quantities captures the amount of velocity flux in the $z$-direction through this cross-section.

We denote the electric field inside a Tokamak by $\mvec{E} + \mvec{E}_{\kappa (t)}^{ind} = (E_{r}, E_{\phi}, E_{z})$. Now, it follows from \eqref{eq:m1} that, $r E_{\phi} = - \int r \frac{\partial B_{z}^{\bar{\chi}}}{\partial t} dr$, where $\mvec{B} + \mvec{B}_{\kappa (t)}^{ind} = B_{r}^{\bar{\chi}} \hat{i}_{r}+B_{\phi}^{\bar{\chi}} \hat{i}_{\phi} + B_{z}^{\bar{\chi}} \hat{i}_{z}$. Let $\kappa$ be regulated by the time-dependent current strength in the outer Tokamak coils, thus, we call it an internal control parameter. It is assumed that the magnetic flux $\int_{S(r)} \mvec{B} \cdot dS$ in the height $z$ direction is increasing with time $t$. This assumption suggests the presence of a strong time-dependent electric field along the toroidal direction $\phi$ inside the Tokamak. Similarly, by increasing the current flow rate through the toroidal coils, we will get a strong time-dependent electric field along the $\phi$ direction. We denote this electric field as $\mvec{E} + \mvec{E}_{\kappa (t)}^{ind}$, following the discussion in \cite{alfven1}, where the authors explained  the existence of an induced electric field $\mvec{E}_{ext} = \frac{E_{0} r_{0}}{r} \hat{i}_{\phi}$, with $r_{0}$ being a reference point. In \cite{alfven1}, however, given that steady state operation was the topic of consideration, it was also assumed that a strong curl-free DC magnetic field exists in the toroidal direction due to the current flow through external poloidal coils. This magnetic field was denoted by $\mvec{B}_{ext} = \frac{B_{0} r_{0}}{r} \hat{i}_{\phi}$. In the present paper, we have assumed that $\bar{\chi}$ depends on time, thereby allowing the operating engineer broad flexibility in selecting a more general consideration of the induced fields, where the $z$-directional magnetic flux is not linearly proportional to time, thereby allowing a time-dependent electric field along the toroidal angular direction $\hat{i}_{\phi}$ with non-zero curl. Moreover, we have kept $B_{0} (t)$ to be a strong DC magnetic field, but also dependent on time. This induced control parameter also controls the velocity of plasma, thereby influencing its shape. 

An important consideration for controlling the Tokamak is that $E_{\phi}$ increases with increasing time in the $\phi$-direction with the charging of current through external coils in Tokamak. We have incorporated the generality regarding the time dependent magnetic field inside the Tokamak with the control of time dependent current flow through external Tokamak coils by incorporating a new control parameter $\kappa (t)$. In the cylindrical coordinates, we can express the external magnetic field as, 
\begin{equation}\label{eq:bcylin}
    \mvec{B} + \mvec{B}_{\kappa (t)}^{ind} = \nabla \bar{\chi} \times \nabla \phi + \left( B_{\phi} + B_{0} (t) \frac{r_{0}}{r} \right) \hat{i}_{\phi}, ~ \bar{\chi} = \chi + \kappa ( t),
\end{equation}
where $\chi, \bar{\chi}$ are two scalar functions (ref. \eqref{eq:chi} for the details of these expressions). The above definition illustrates the fact that, $\kappa (t) $, called internal control parameter, is added in the magnetic flux term $\chi = \frac{1}{2 \pi} \int \mvec{B} \cdot ds$ to incorporate the enhancement in magnetic field due to a rise in external time dependent current intensity, considered as an external control parameter. By doing further calculations, we can derive that the $\phi$ component of the electric field 
\begin{equation}\label{eq:ecylinphi}
[\mvec{E} + \mvec{E}_{\kappa (t)}^{ind}]_{\phi} = \frac{1}{r} \left( \frac{\partial \chi}{\partial t} + \kappa^{'} (t) \right).  
\end{equation}
As evident from these equations, the reason for calling the term $\kappa$ an ``intrinsic control parameter" lies with the fact the effects of $\kappa$ are visible in the electric field $\mvec{E} + \mvec{E}_{\kappa (t)}^{ind}$ without altering the behavior of $\mvec{B} + \mvec{B}_{\kappa (t)}^{ind}$ in the sense that $\nabla \bar{\chi} = \nabla \chi$ (ref. \eqref{eq:chi}). The equations \eqref{eq:max2}, \eqref{eq:div} represent the plasma flow equations in the presence of an external body force, namely the Lorentz force. The equations \eqref{eq:m1}-\eqref{eq:max3} represent the Faraday's law, Ohm' law and Amp\`{e}re's law respectively. 

In the initial discharge phase, the current density through toroidal coils are gradually increased to achieve a steady-state magnetic field that would contain plasma inside a Tokamak without touching its boundary. Subsequently, the applied current density through external coils can act as a feedback control parameter that would send a signal to control the intensity of applied current density in plasma and the magnetic field inside the Tokamak, thereby maintaining a steady and stable plasma inside the device. Thus, we have introduced two control parameters $\kappa (t)$ and $B_{0} (t)$ to incorporate the effects of both toroidal and poloidal currents through external coils on the magnetic field inside the Tokamak.
Another way of controlling the current is by neutral beam injection \cite{beam, beamarticle}; however, the symmetric cylindrical conditions for modeling a Tokamak would have to be dropped to consider such a process. However, one approach towards mathematically modeling this operation is presented in \cite{OriginalEq}, where the modified Navier-Stokes equations, due to Olga Ladyzhenskaya \cite{Ladyzhenskaya1998, Lady2}, are used.

We denote the canonical definition of a vector field in the cylindrical coordinate system as well as a differential operator we shall use in the sequel:
\begin{align*}
    \nabla f(r, \phi, z) &= \frac{\partial f}{\partial r} \hat{i}_{r} + \frac{1}{r} \frac{\partial f}{\partial \phi} \hat{i}_{\phi} + \frac{\partial f}{\partial z} \hat{i}_{z}, \nonumber \\
{\Delta}^{*} A = \nabla^{2} A - \frac{2}{r} \frac{\partial A}{\partial r} 
               &= \frac{\partial^2 A}{\partial r^2} - \frac{1}{r} \frac{\partial A}{\partial r} + \frac{\partial^2 A}{\partial z^2}, \nonumber 
\end{align*}
In addition, we introduce two intermediate quantities, the plasma current and the vorticity as,
\begin{align*} 
  \mvec{J}_{p}:=\nabla \times \left(\mvec{B} + \mvec{B}_{\kappa (t)}^{ind} \right) , 
    &\qquad 
    \mvec{\omega}:=\nabla \times \mvec{v}. \nonumber
 \end{align*}
We use the following substitutions (appearing in \cite{alfven1, alfven2, alfven3, Marco}) to express the model of the unsteady electromagnetic plasma inside the Tokamak in cylindrical coordinates, 
\begin{align} 
\mvec{v}(r,z) &= \nabla \psi \times \nabla \phi + v_{\phi} \hat{i}_{\phi}, 
       \qquad \qquad \qquad \text{$\nabla \phi = r^{-1} \hat{i}_{\phi}$},
\label{eq:alfven}
\\
\mvec{B}(r,z) + \mvec{B}_{\kappa (t)}^{ind} &= \nabla \bar{\chi} \times \nabla \phi +  \left(B_{\phi} + B_{0} (t) \frac{r_0}{r} \right) \hat{i}_{\phi}, 
\label{eq:alfvenmagnet}
\\
\mvec{J}_{p}(r,z) &= \nabla (r B_{\phi}) \times \nabla \phi - \frac{1}{r} ({\Delta}^{*} \bar{\chi}) \hat{i}_{\phi},
\label{eq:alfvencurrent} 
\\
\mvec{\omega}(r,z)  &= \nabla (r v_{\phi}) \times \nabla \phi - \frac{1}{r} ({\Delta}^{*} \psi) \hat{i}_{\phi},  
\label{eq:alfvenomega} 
\end{align}
The calculations deriving these expressions are given in Appendix \ref{sec:appcalc}. 

While these equations represent a geometrically transparent picture of the MHD dynamics, we shall introduce a set of new coordinates and variable components to facilitate amenable computation. This appears in the literature and incorporates Tokamak specific symmetries and presents the system in a manner suitable for engineering considerations. 
\begin{align} \label{eq:alfven2} 
  a    & := \frac{r}{r_0}, 
~ b      := \frac{z}{r_0},
~ u_{1}  := \frac{\psi}{r_0},
~ u_{2}  := r_{0} r \omega_{\phi},
~ u_{3}  := r B_{\phi},
\nonumber \\
~ u_{4} &:= r v_{\phi},
~  u_{5}  := \frac{\chi}{r_0},
~ u_{6}  := r_{0} r {J_{p}}_{\phi},
~ A          := \frac{\chi_{0}}{r_{0}}, 
\end{align}
where $\chi_{0}$ is a constant scalar function and $\eta := \frac{1}{\sigma_{p}}$, where $\sigma_{p}$ denotes the electrical conductivity of plasma.
Here, the maximum height of the Tokamak is denoted by $z_{0} r_{0}$ and maximum radial length is denoted by $r_{m} r_{0}$. By assuming $r_{0}$ to be the middle point in the radial direction, $r_{l} r_{0}$ the left most point of Tokamak in the radial direction, for the rest of this paper, we denote the cross-sectional region $[r_{l} r_{0}, r_{m} r_{0}] \times [0, 2 \pi] \times [-z_{0} r_{0}, z_{0} r_{0}] $ by the symbol $\mcdomain$. See Figure~\ref{Diagram} for an illustration. In the sequel we shall denote the set of spatial coordinates as $x=(a,b,\phi)$

With these new set of variables, we write the set of PDEs governing the plasma
flow as follows,
\begin{align} \label{eq:reduced}
\overline{{\Delta}^{*}} u_{1} & = - u_{2}, \\
\nu \overline{\Delta^{*}} u_{2} &= r_{0}^{2} \frac{\partial u_{2}}{\partial t} + \frac{1}{a} \{ u_{1}, u_{2} \} + \frac{2}{a^{2}} u_{2} \frac{\partial u_{1}}{\partial b} + \frac{1}{a} \{ u_{6}, u_{5} \} - \frac{2}{a^{2}} u_{6} \frac{\partial u_{5}}{\partial b}  \\
    & \quad + \frac{2}{a^{2}} \Big[ \left(u_{3} + r_{0} B_{0} (t) \right) \frac{\partial u_{3}}{\partial b} - u_{4} \frac{\partial u_{4}}{\partial b} \Big], \\
    \eta \overline{\Delta^{*}} u_{3} &= r_{0}^{2} \frac{\partial u_{3}}{\partial t} + r_{0}^{3} B_{0}^{'}(t) + \frac{1}{a} \Big( \{ u_{4}, u_{5} \} + \{ u_{1}, u_{3} \} \Big) \nonumber \\
    & \quad + \frac{2}{a^{2}} \Big( \left(u_{3} + r_{0}B_{0} (t) \right) \frac{\partial u_{1}}{\partial b} - u_{4} \frac{\partial u_{5}}{\partial b} \Big),   \\
\nu \overline{\Delta^{*}} u_{4} &= r_{0}^{2} \frac{\partial u_{4}}{\partial t} + \frac{1}{a} \Big( \{ u_{3}, u_{5} \} + \{ u_{1}, u_{4} \} \Big), \\
    \overline{{\Delta}^{*}} u_{5} &= - u_{6} , \\ 
\eta u_{6} &= r_{0}^{2} \frac{\partial u_{5}}{\partial t} + r_{0} \kappa^{'} (t) + \frac{1}{a} \{ u_{5}, u_{1} \},
\label{eq:control1}
\end{align}
where ${\Delta}^{*}$ in terms of $a, b$ takes the form
$\overline{{\Delta}^{*}} = \frac{\partial^2}{\partial a^2} + \frac{\partial^2}{\partial b^2} - \frac{1}{a} \frac{\partial}{\partial a} $ all of the equations are defined to be on the domain $\mcdomain\times[0,T]$. See Appendix~\ref{sec:apptran} for the derivation of this system.

These equations are completed with the set of boundary conditions which are taken from \eqref{eq:rectbound}, transformed into cylindrical coordinates (see Appendix Section~\ref{sec:appcylbound}) and finally expressed below in the new set of coordinates $\{u_1, u_2, u_{3}, u_{4}, u_5, u_{6}\}$ in equations \eqref{eq:boundaryb1}-\eqref{eq:boundaryb5}, the calculations for which are derived in Section~\ref{sec:apptranbound},
\begin{align}
    \frac{\partial u_{5}}{\partial b} &= 0, & \text{On } \partial \mcdomain \times [0,T] \label{eq:boundaryb1} \\
    \frac{\partial u_{3}}{\partial a} &= 0, & \text{On } \partial \mcdomain \times [0,T] \\
    \Big[\frac{\partial^{2} u_{5}}{\partial a^{2}} - \frac{1}{a} \frac{\partial u_{5}}{\partial a} \Big] &= 0, & \text{On } \partial \mcdomain \times [0,T] \label{eq:boundaryb2} \\
    u_{1} &= 0, & \text{On } \partial \mcdomain \times [0,T] \\
    u_{4} &= 0,  & \text{On } \partial \mcdomain \times [0,T] \label{eq:boundaryb3}\\
    u_{1}(0) &= 0, & \text{on } \mcdomain  \\
    u_{4}(0) &= 0, & \text{on } \mcdomain  \label{eq:boundaryb4}\\
    u_{5}(0) &= A, & \text{on } \mcdomain  \\
    u_{3}(0) &= a r_{0} B_{1} - r_{0} B_{0}(0), & \text{on } \mcdomain  \label{eq:boundaryb5}
\end{align}
where $B_1 \in \mathbb{R}$ is a real number. The boundary conditions indicate that there is no current at the outer surface of the Tokamak, which is necessary for its safe operation and durability. The initial conditions indicate the operation is at an initial static start up. 


\subsection{Objective Function}
In this paper, we are interested in controlling the shape of plasma and induced magnetic field lines inside the Tokamak as expressed by its velocity field and induced magnetic field, respectively. We have two target functions $\mvec{B}_d(\cdot,T)$ and $\mvec{v}_d(\cdot,T)$ that correspond at time $T$ to a desired, predetermined steady state velocity and magnetic fields respectively. We assume these have been determined a priori as the dynamic of a steady state optimal system. The control parameter for our problem is an internal control parameter in plasma $\kappa (t)$ and an external control parameter $B_{0} (t)$. 

Now we are ready to present our optimal control problem in the cylindrical coordinate system. For a fixed time parameter $T>0$, we are interested to minimize the following cost functional,
\begin{align} \label{eq:costf}
J(\mvec{v}, \mvec{B}, \kappa (t), B_{0} (t)) &=   \frac{\alpha_1}{2} \|\mvec{v}(\cdot ,T) - \mvec{v_d}(\cdot ,T)\|^{2} \nonumber \\
& \quad + \frac{\alpha_2}{2} \|\mvec{B} + \mvec{B}_{\kappa (t)}^{ind}(\cdot ,T) - \mvec{B}_{0}(T) - \mvec{B_d}(\cdot , T)\|^{2}  \nonumber \\
& \quad + \int_{0}^{T} \left(\frac{\beta_1}{2} \kappa^{2}(t)  + \frac{\beta_2}{2} {B_{0}}^{2}(t) \right) dt  , \nonumber  \\
&= \frac{\alpha_1}{2 a^{2}} \left\|- \frac{\partial u_{1}}{\partial b} \mvec{\hat{i}}_{r} + \frac{\partial u_{1}}{\partial a} \mvec{\hat{i}}_{z} +  u_{4} \mvec{\hat{i}}_{\phi} (\cdot ,T) - a r_{0} \mvec{v_d}(\cdot ,T) \right\|^{2} \nonumber \\ 
& \quad + \frac{\alpha_2}{2 a^{2}} \left\|- \frac{\partial u_{5}}{\partial b} \mvec{\hat{i}}_{r} + \frac{\partial u_{5}}{\partial a} \mvec{\hat{i}}_{z} + u_{3} \mvec{\hat{i}}_{\phi} (\cdot ,T) - a r_{0} \mvec{B_d}(\cdot , T) \right\|^{2}  \nonumber \\
& \quad + \int_{0}^{T} \left(\frac{\beta_1}{2} \kappa^{2}(t)  + \frac{\beta_2}{2} {B_{0}}^{2} (t) \right)  r_{0}^{2} dt , \nonumber \\
&= \int_{r_{l}}^{r_{m}} \int_{-z_{0} }^{z_{0} } \Bigg[ \frac{\alpha_1}{2 \bar{a}^{2}} \Big[ \left( \frac{\partial u_{1}}{\partial b} + \bar{a} r_{0} {v_{d}}_r \right)^{2} + \left( \frac{\partial u_{1}}{\partial a} - \bar{a} r_{0} {v_{d}}_z \right)^{2} \nonumber \\
& \quad + \left( u_{4} - \bar{a} r_{0} {v_{d}}_{\phi} \right)^{2} \Big] + \frac{\alpha_2}{2 \bar{a}^{2}} \Big[ \left( \frac{\partial u_{5}}{\partial b} + \bar{a} r_{0} [B_{d}]_r \right)^{2} \nonumber \\
& \quad + \left( \frac{\partial u_{5}}{\partial a} - \bar{a} r_{0} [B_{d}]_z \right)^{2} + \left( u_{3} - \bar{a} r_{0} [B_{d}]_{\phi} \right)^{2} \Big]\Bigg] d\bar{a} ~ d\bar{b} \nonumber \\
& \quad +  \int_{0}^{T} \left( \frac{\beta_1}{2} \kappa^{2}(t)  + \frac{\beta_2}{2} {B_{0}}^{2} (t) \right)  r_{0}^{2} dt  
\end{align}
where $\mvec{v}_{d} = v_{d}^{1} \hat{i}_{r} + v_{d}^{2} \hat{i}_{z} + v_{d}^{3} \hat{i}_{\phi}, ~ \mvec{B}_{d} = B_{d}^{1} \hat{i}_{r} + B_{d}^{2} \hat{i}_{z} + B_{d}^{3} \hat{i}_{\phi}, ~ \mvec{B}_{0} (T) = B_{0} (T) \frac{r_{0}}{r} \hat{i}_{\phi}$, subject to \eqref{eq:alfven}, \eqref{eq:alfvenmagnet}, \eqref{eq:reduced} and \eqref{eq:boundaryb1}-\eqref{eq:boundaryb4}, for target desired steady state flow $v_d$ and magnetic field $B_d$ and we regularize the controls to ensure a well defined solution.
The constants $\alpha_{1}, \alpha_{2}, \beta_{1},\beta_2 \ge 0$ denote the non negative weights on the quantities of our interest. $\alpha_1>0$ indicates a desired velocity field, $\alpha_2>0$ a desired magnetic field, and typically $\beta_1,\beta_2>0$ but small, serving as a regularization to encourage unique solutions. 

\section{Robust Optimal Control Problem} \label{stochastic} 
It is found in the literature \cite{AGREDANOTORRES2021112683} that the electrical conductivity of plasma changes with plasma temperature.
As plasma is kept at a very high temperature inside a Tokamak, which can then moderately vary in an unpredictable manner, it is naturally concerning to assume the electrical conductivity of plasma $\sigma_{p}$ to be a fixed and known quantity.
Moreover, the uncertainty of $\sigma_{p}$ is an important determining factor in heat generation inside the Tokamak by Ohmic heating, thereby disregarding its uncertainty could yield improper operation (see, e.g. \cite{Marco}).
In this paper, the electrical conductivity of plasma $\sigma_{p}$ and kinematic viscosity of plasma $\nu$ are considered to be two stochastic coefficients. We amend the optimal control problem to be robust with respect to the uncertainty, by targeting that the expected deviation from the target is minimized. For a background on stochastic PDE constrained optimization, see~\cite{kouri2018optimization}. We note that in lieu of the expectation, we could use risk measures as a means of particularly penalizing, e.g., far from steady state, or in some sense physically hazardous velocity and magnetic fields, however, we stick to the expectation as a simpler first presentation in the literature of the consideration of uncertainty in the operation of a Tokamak.

\subsection{Stochastic PDE Systems}
We assume $\sigma_{p}, \nu$ to be two second order random fields such that $\sigma_{p}, \nu \in L^{2} (\mcdomain) \otimes L^{2} (\Xi) $, where $(\Xi,\mathcal{F},\mathcal{P})$ denotes a complete probability space with $\Xi$ representing a sample space, $\mathcal{F}$ a $\sigma$-algebra with a probability measure $\mathcal{P}$.
This consideration of $\sigma_{p} (\xi), \nu(\xi)$ as two random fields automatically makes the state variables $\mvec{v}$ and $\mvec{B}$ random quantities as well.
These two second-order random fields $\mvec{v}$ and $\mvec{B}$ again belong to a tensor product Hilbert space $L^{2} (\mcdomain) \otimes L^{2} (\Xi)$, which is defined below,
\begin{align} 
    L^{2} (\mcdomain) \otimes L^{2} (\Xi) = \{ \omega (x, \xi) \colon \mcdomain \otimes \Xi \to  \mathbb{R} \colon \int_{\Xi} \int_{\mcdomain} |\omega|^{2} ds ~ d \mathcal{P} < \infty \}, \nonumber
\end{align}
with the norm,
\begin{align} 
    \|\omega\|_{L^{2} (\mcdomain) \otimes L^{2} (\Xi)} = \Big ( \int_{\mcdomain} \int_{\Xi} 
|\omega|^{2} ds ~ d \mathcal{P} \Big )^{\frac{1}{2}}. \nonumber
\end{align}
With the above notations at our hand, we modify the PDE systems \eqref{eq:reduced}-\eqref{eq:control1} along with the constraint boundary conditions \eqref{eq:boundaryb1}-\eqref{eq:boundaryb5} into a minimization problem with stochastic PDE constraint equations as described below,
\begin{align}
    \min_{v(\cdot,\xi), B(\cdot,\xi), B_{0} (t), \kappa (t)} \mathbb{E}_{\xi}[J(\mvec{v},\mvec{B},\kappa(t),B_0(t))]
\end{align}
subject to:
\begin{align*} 
\overline{{\Delta}^{*}} u_{1}(\xi) & = - u_{2}(\xi), \\
\nu \overline{\Delta^{*}} u_{2}(\xi) &= r_{0}^{2} \frac{\partial u_{2}(\xi)}{\partial t} + \frac{1}{a} \{ u_{1}, u_{2} \}(\xi) + \frac{2}{a^{2}} u_{2}(\xi) \frac{\partial u_{1}(\xi)}{\partial b} + \frac{1}{a} \{ u_{6} , u_{5} \}(\xi)  \\
    & \quad - \frac{2}{a^{2}} u_{6}(\xi)  \frac{\partial u_{5}(\xi)}{\partial b} + \frac{2}{a^{2}} \Big[ \left(u_{3}(\xi) + r_{0} B_{0} (t) \right) \frac{\partial u_{3}(\xi)}{\partial b} - u_{4}(\xi) \frac{\partial u_{4}(\xi)}{\partial b} \Big], \\
    \eta \overline{\Delta^{*}} u_{3}(\xi) &= r_{0}^{2} \frac{\partial u_{3}(\xi)}{\partial t} + r_{0}^{3} B_{0}^{'} (t) + \frac{1}{a} \Big( \{ u_{4}, u_{5} \}(\xi) + \{ u_{1}, u_{3} \}(\xi) \Big) \\
    & \quad + \frac{2}{a^{2}} \Big( \left( u_{3}(\xi) + r_{0} B_{0} (t) \right) \frac{\partial u_{1}(\xi)}{\partial b}  - u_{4}(\xi) \frac{\partial u_{5}(\xi)}{\partial b} \Big),   \\
\nu \overline{\Delta^{*}} u_{4}(\xi) &= r_{0}^{2} \frac{\partial u_{4}(\xi)}{\partial t} + \frac{1}{a} \Big( \{ u_{3}, u_{5} \}(\xi) + \{ u_{1}, u_{4} \}(\xi) \Big), \\
    \overline{{\Delta}^{*}} u_{5} (\xi) &= - u_{6}(\xi), \\ 
\eta u_{6}(\xi) &= r_{0}^{2} \frac{\partial u_{5}(\xi)}{\partial t} + r_{0} \kappa^{'} (t) + \frac{1}{a} \{ u_{5}, u_{1} \}(\xi)
\end{align*}
and boundary conditions:
\begin{align}
    \frac{\partial u_{5}}{\partial b} (x, t, \xi) &= 0, & \text{on }\partial \mcdomain \times [0,T] \nonumber \\
    \frac{\partial u_{3}}{\partial a} (x, t, \xi) &= 0, & \text{on }\partial \mcdomain \times [0,T]\nonumber \\
    \Big[\frac{\partial^{2} u_{5}}{\partial a^{2}} - \frac{1}{a} \frac{\partial u_{5}}{\partial a} \Big] (x, t, \xi)  &= 0, & \text{on }\partial \mcdomain \times [0,T] \nonumber \\
    u_{1} (x, t, \xi) &= 0, & \text{on }\partial \mcdomain \times [0,T] \nonumber \\
    u_{4} (x, t, \xi) &= 0, & \text{on }\partial \mcdomain \times [0,T] \nonumber \\
    u_{1} (x,0,\xi) &= 0, & \text{on } \mcdomain  \nonumber \\
   u_{4} (x,0,\xi) &= 0, & \text{on } \mcdomain  \nonumber \\
   u_{5} (x,0,\xi) &= A, & \text{on } \mcdomain  \nonumber \\
   u_{3} (x,0,\xi) &= a r_{0} B_{1} - r_{0} B_{0} (0), & \text{on } \mcdomain \nonumber
\end{align}
and with cost functional:
\begin{align} 
\mathcal{J}_{1} &= \frac{\alpha_{1}}{2} \|\mvec{v} (\cdot , T) - \mvec{v}_{d} (\cdot , T)\|_{L^{2}(\mcdomain) \otimes L^{2} (\Xi)}^{2} + \frac{\alpha_{2}}{2} \|\mvec{B} (\cdot , T) - \mvec{B}_{0} (T) - \mvec{B}_{d} (\cdot , T)\|_{L^{2}(\mcdomain) \otimes L^{2} (\Xi)}^{2} \nonumber \\
& \quad
+ \int_{0}^{T} \left[ \frac{\beta_{1}}{2} (B_{0} (t))^{2} 
+ \frac{\beta_{2}}{2} (\kappa (t))^{2} \right] dt ,
\label{eq:20}
\end{align}
Here, the newly defined cost function $\mathcal{J}_{1}$ in equation \eqref{eq:20} is the expectation value of the stochastic functional $J$, thereby producing a deterministic outcome. 

\subsection{Finite dimensional representation of random fields}
We will employ the finite-dimensional noise assumption to describe $\sigma_{p}$ which states that $\sigma_{p}$ can be approximated by a finite number of random variables $\xi = \{\xi_{i} \}_{t = 1}^{L}$, for a natural number $L$ and $\xi_{i} \colon \Xi \to  \Gamma_{i} \subset \mathbb{R}$.  Additionally, we assume that $\xi_{i}$ are independent random variables having a probability density function $\rho_{i} \colon \Gamma_{i} \to  [0,1]$. The joint probability distribution of $\xi$ is expressed as $\rho = \Pi_{i = 1}^{L} \rho_{i} (y_{i})$ where $(y_{1}, \cdots, y_{L}) \in \Gamma = \Pi_{i = 1}^{L} \Gamma_{i} \subset \mathbb{R}^{L}$. The truncated Karhunen-Lo\`{e}ve expansion (see, e.g.,~\cite{wang2008karhunen}) of $\sigma_{p}$ and $\nu$ is expressed as,
\begin{align}
      \sigma_{p} (x, y) &= \sum_{i = 1}^{L + 1} {\sigma_{p}}_{i} (x) \zeta_{i} (y), &\text{On }x \in \mcdomain, ~ y \in \Gamma,                 \nonumber \\
      \nu (x, y) &= \sum_{i = 1}^{L + 1} {\nu}_{i} (x) \zeta_{i} (y) &\text{On }x \in \mcdomain, ~ y \in \Gamma, \nonumber
\end{align}
where ${\sigma_{p}}_{i}, \nu_{i} \colon \mcdomain \to  \mathbb{R}$ and $\zeta_{i} = y_{i - 1}$ with $y_{0} = 1$. 
For the two separable functions 
$u_{1}, u_{2} \colon \mcdomain \times \Gamma \to  \mathbb{R}$, 
we define an inner product between them using the individual components
$v_{1}, v_{2} \in H^{1}(\mcdomain)$
and
$w_{1}, w_{2} \in L_{\rho}^{2}(\Gamma)$
of their separable representations:
\begin{align*}
    u_{1} (x,y) = v_{1} (x) w_{1} (y), 
    \quad
    u_{2} (x, y) &= v_{2} (x) w_{2} (y),
\end{align*}
through the following inner-product definition:
\begin{align*}
(u_{1}, u_{2})_{H^{1} (\mcdomain) \otimes H_{\rho}^{1} (\Gamma)} &= (v_{1}, v_{2})_{H^{1} (\mcdomain)} (w_{1}, w_{2})_{H_{\rho}^{1} (\Gamma)},
\end{align*}
where
\begin{align*}
    (v,w)_{L_{\rho}^{2} (\Gamma)} &= \int_{\Gamma} v w \rho dy\text{ for }v,w \in L_{\rho}^{2} (\Gamma).
\end{align*}
Finally, using the Doob-Dynkin Lemma, the random vectors $\mvec{v}, \mvec{B}, \sigma_{p}$ can be expressed as a function of $L$ random variables, thereby allowing us to reformulate the stochastic optimal control problem, defined earlier in \eqref{eq:20}, in the following manner,
\begin{align}
    \min_{\mvec{v}, \mvec{B}, B_{0} (t), \kappa (t)} \mathcal{J}_{1} \nonumber
\end{align}
where:
\begin{align} \label{eq:stochastic}
\mathcal{J}_{1} &= \frac{\alpha_{1}}{2} ||\mvec{v} - \mvec{v}_{d}||_{L^{2}(\mcdomain) \otimes L_{\rho}^{2} (\Gamma)}^{2} + \frac{\alpha_{2}}{2} ||\mvec{B} - \mvec{B}_{0} (T) - \mvec{B}_{d}||_{L^{2}(\mcdomain) \otimes L_{\rho}^{2} (\Gamma)}^{2}  \nonumber \\
& \quad + \int_{0}^{T} \Big[  \frac{\beta_{1}}{2} (B_{0} (t))^{2} + \frac{\beta_{2}}{2} (\kappa(t))^{2} \Big] ~ ~ dt, 
\end{align}

\subsection{Lagrangian}
Introduce the set of adjoint multipliers, 
\[
\lambda_{1}, \lambda_{2}, \lambda_{3}, \lambda_{4}, \lambda_{5}, \mu_{1}, \mu_{2}, \mu_{3}, \mu_{4}, \mu_{5}, \mu_{6} \in H_{0}^{1} (\mcdomain) \otimes L^{2}_{\rho} (\Gamma)
\]
to construct a Lagrangian $\mathcal{L}_{1}$ associated with the stochastic cost function as follows,
\begin{align} \label{eq:lagrangian}
    \mathcal{L}_{1} &=  \int_{\mcdomain}  \int_{\Gamma} \Bigg[ \frac{\alpha_1}{2 a^{2}} \Big[ \left( \frac{\partial u_{1}}{\partial b} + a r_{0} [v_{d}]_{r} \right)^{2} + \left( \frac{\partial u_{1}}{\partial a} - a r_{0} [v_{d}]_{z} \right)^{2} \nonumber \\
& \quad + \left( u_{4} - a r_{0} [v_{d}]_{\phi} \right)^{2} \Big] + \frac{\alpha_2}{2 a^{2}} \Big[ \left( \frac{\partial u_{5}}{\partial b} + a r_{0} [B_{d}]_{r} \right)^{2} + \left( \frac{\partial u_{5}}{\partial a} - a r_{0} [B_{d}]_z \right)^{2} \nonumber \\
& \quad + \left( u_{3} - a r_{0} [B_{d}]_{\phi} \right)^{2} \Big] \Bigg] \rho da  db   dy +  \int_{0}^{T}  \left( \frac{\beta_1}{2} \kappa^{2} (t)  + \frac{\beta_2}{2} {B}_{0}^{2} (t) \right)  r_{0}^{2} dt    \nonumber \\
    & \quad-  \int_{0}^{T} \int_{\partial \mcdomain}  \int_{\Gamma} \lambda_{1} \frac{\partial u_{5}}{\partial b} ~ \rho  da  db  dy  dt - \int_{0}^{T} \oint_{\partial C}  \int_{\Gamma} \lambda_{2} \frac{\partial u_{3}}{\partial a} ~ \rho  da  db  dy  dt \nonumber \\
    & \quad- \int_{0}^{T} \int_{\partial \mcdomain}  \int_{\Gamma} \lambda_{3} \Big[ \frac{\partial^{2} u_{5}}{\partial a^{2}} - \frac{1}{a} \frac{\partial u_{5}}{\partial a} \Big] ~ \rho dy  da  db  dt \nonumber \\
& \quad- \int_{0}^{T} \int_{\partial \mcdomain} \int_{\Gamma}
 (\lambda_{4} u_{1} + \lambda_{5} u_{4}) ~ \rho  da  db  dy  dt - \int_{0}^{T} \int_{\mcdomain} \int_{\Gamma} \mu_{1} (\overline{{\Delta}^{*}} u_{1} + u_{2}) ~ \rho dy da db dt \nonumber \\
& \quad- \int_{0}^{T} \int_{\mcdomain} \int_{\Gamma} \mu_{2} \Big[ - \nu \overline{\Delta^{*}} u_{2} + r_{0}^{2} \frac{\partial u_{2}}{\partial t} + \frac{1}{a} \{ u_{1}, u_{2} \} + \frac{2}{a^{2}} u_{2} \frac{\partial u_{1}}{\partial b} + \frac{1}{a} \{ u_{6} , u_{5} \} \nonumber \\
& \quad - \frac{2}{a^{2}} u_{6} \frac{\partial u_{5}}{\partial b}  + \frac{2}{a^{2}} \Big[ \left( u_{3} + r_{0} B_{0} (t) \right) \frac{\partial u_{3}}{\partial b} - u_{4} \frac{\partial u_{4}}{\partial b} \Big] \Big] ~ \rho da  db  dy  dt \nonumber  \\
& \quad- \int_{0}^{T} \int_{\mcdomain} \int_{\Gamma} \mu_{3} \Big[ - \eta \overline{\Delta^{*}} u_{3} + r_{0}^{2} \frac{\partial u_{3}}{\partial t} + r_{0}^{3} B_{0}^{'} (t) + \frac{1}{a} \Big( \{ u_{4}, u_{5} \} + \{ u_{1}, u_{3} \} \Big) \nonumber \\
& \quad + \frac{2}{a^{2}} \Big( \left( u_{3} + r_{0} B_{0} (t) \right)\frac{\partial u_{1}}{\partial b}  - u_{4} \frac{\partial u_{5}}{\partial b} \Big) \Big] ~ \rho  da  db  dy  dt \nonumber \\
& \quad- \int_{0}^{T} \int_{\mcdomain} \int_{\Gamma} \mu_{4} 
\Big[ - \nu \overline{\Delta^{*}} u_{4} + r_{0}^{2} \frac{\partial u_{4}}{\partial t} + \frac{1}{a} \Big( \{ u_{3}, u_{5} \} + \{ u_{1}, u_{4} \} \Big) \Big] ~ \rho  da  db  dy  dt \nonumber \\
& \quad- \int_{0}^{T} \int_{\mcdomain} \int_{\Gamma} \mu_{5} (
\overline{{\Delta}^{*}} u_{5} + u_{6}) ~ \rho  da  db  dy  dt \nonumber \\ 
& \quad- \int_{0}^{T} \int_{\mcdomain} \int_{\Gamma} \mu_{6} (\eta u_{6} - r_{0}^{2} \frac{\partial u_{5}}{\partial t} - r_{0} \kappa^{'} (t) - \frac{1}{a} \{ u_{5}, u_{1} \}) ~ \rho  da  db  dy  dt ,  \end{align}
where the boundary conditions $u_{1 , 4}(\cdot, 0) = 0, ~ u_{5}(\cdot,0) = A, ~ u_{3}(\cdot,0) = a r_{0} B_{1} - r_{0} B_{0} (0)$ must be imposed by construction via the definition of the function spaces wherein $\mvec{v}, \mvec{B}$ lie. 

\subsection{Necessary Optimality Conditions}
We are interested in finding the stationary points of the Lagrangian \eqref{eq:lagrangian}, with respect to the adjoint, state and control variables in the sense of variations. By taking the directional derivatives of $\mathcal{L}_{1}$ with respect to $\lambda_{i}, \mu_{j}$ and setting it equal to zero for all possible directions, we obtain the constraint conditions \eqref{eq:reduced}, \eqref{eq:boundaryb1}-\eqref{eq:boundaryb5}. The rest of the optimality conditions characterizing a stationary solution to the optimal control problem are derived by taking variational derivatives with respect to the states $\{u_i\}$ and controls $\kappa (t)$ and $B_{0} (t)$. 

\noindent It holds that $D_{\kappa (t)}\mathcal{L}_{1} \bar{f}=0$ if and only if for all $f \in L^2([0,T])$,
\begin{align*}
& \int_0^T\beta_1 r_{0}^{2} \kappa (t) \bar{f} f dt -\int_0^T\int_{\mcdomain}\int_{\Gamma}  \mu_6^{'} (t) r_0  \rho dydadb\bar{f} f dt = 0 \\
&\iff \\
& \beta_1 r_0 \kappa (t) - \mathbb{E}\left[\int_{\mcdomain}\mu_6^{'} (\cdot,t)\rho dadb\right] = 0 \text{ on }[0,T]
\end{align*}
Note that this implies that the reduced gradient is,
\begin{align}
    \frac{d \hat{J}_1}{d \kappa (t)} = \beta_1 r_0 \kappa (t) -  \mathbb{E}\left[\int_{\mcdomain}\mu_6^{'} (\cdot,t)\rho dadb\right] .
\end{align}
It holds that $D_{B_{0} (t)}\mathcal{L}_{1} {\bar{B}}=0$ if and only if for all $B \in L^2([0,T])$,
\begin{align*}
& \int_0^T\beta_2 r_{0}^{2} B_{0} (t) {\bar{B}} Bdt - \int_0^T\int_{\mcdomain}\int_{\Gamma}  \frac{2 r_{0}
}{a^{2}} \mu_{3} \frac{\partial u_{3}}{\partial b}  \rho dydadb{\bar{B}} Bdt \\
& - \int_0^T\int_{\mcdomain}\int_{\Gamma}  \frac{2 r_{0} \mu_3}{a^2} \frac{\partial u_1}{\partial b}  \rho dydadb{\bar{B}} Bdt  + \int_0^T\int_{\mcdomain}\int_{\Gamma}  r_{0}^{3}  \mu_{3}^{'} (t) \rho dydadb{\bar{B}} Bdt \\
&\iff \\
& \beta_{2} r_{0}^{2} B_{0} (t) + \mathbb{E}\left[\int_{\mcdomain}r_{0}^{3}  \mu_{3}^{'} (\cdot,t) \rho dadb\right] - \mathbb{E} \left[\int_{\mcdomain}\frac{2 r_{0}}{a^{2}} \mu_{3}(\cdot,t) \frac{\partial u_{3}}{\partial b} \rho dadb\right] \\ &\qquad \qquad - \mathbb{E} \left[ \int_{\mcdomain}\frac{2 r_{0} \mu_3}{a^2} \frac{\partial u_1}{\partial b} \rho dadb\right] = 0 \text{ on }[0,T]
\end{align*}
Note that this implies that the reduced gradient is,
\begin{align*}
    \frac{d\hat{J}_1}{d B_0} = & \beta_{2} r_{0}^{2} B_{0} (t) + \mathbb{E}\left[\int_{\mcdomain}r_{0}^{3}  \mu_{3}^{'} (\cdot,t) \rho dadb\right] - \mathbb{E} \left[\int_{\mcdomain}\frac{2 r_{0}}{a^{2}} \mu_{3}(\cdot,t) \frac{\partial u_{3}}{\partial b} \rho dadb\right] \\ &\qquad \qquad - \mathbb{E} \left[ \int_{\mcdomain}\frac{2 r_{0} \mu_3}{a^2} \frac{\partial u_1}{\partial b} \rho dadb\right].
\end{align*}
The variational derivative of $\mathcal{L}_{1}$ with respect to the state variable $u_{6}$ in the direction $\bar{u} \in L^2([0,T],L^2(\mcdomain))\otimes L_{\rho}^{2} (\Gamma)$ reads:
\begin{align}
    D_{u_{6}} \mathcal{L}_{1} \bar{u} = \int_{0}^{T} \int_{\mcdomain} \int_{\Gamma} \left[ \frac{2}{a^2} \mu_{2} \frac{\partial u_{5}}{\partial b} - \mu_{5} - \eta \mu_{6} + \frac{1}{a} \{ \mu_2, u_5 \} \right] \bar{u} \rho dy da db dt \nonumber
\end{align}
Now it holds that $D_{u_{6}} \mathcal{L}_{1} \bar{u} = 0$ if and only if for all $w \in L^2([0,T],L^2(\mcdomain)) \otimes L_{\rho}^{2} (\Gamma)$,
\begin{align}
    \frac{2}{a^2} \mu_{2} \frac{\partial u_{5}}{\partial b} - \mu_{5} - \eta \mu_{6} + \frac{1}{a} \{ \mu_2, u_5 \} = 0.
\end{align}
We now calculate the following,
\begin{align}
    D_{u_{2}} \mathcal{L}_{1} \bar{u} = - \int_{0}^{T} \int_{C} \int_{\Gamma} \left[ \mu_{1} - r_{0}^{2}  \frac{\partial \mu_{2}}{\partial t} + \nu \overline{\Delta^{*}} \mu_{2} - \frac{1}{a} \{ u_1, \mu_2 \} \right] \bar{u} \rho dy da db dt \nonumber
\end{align}
Now it holds that $D_{u_{2}} \mathcal{L}_{1} \bar{u} = 0$ if and only if for all $w \in H^1([0,T],H^1(\mcdomain)) \otimes L_{\rho}^{2} (\Gamma)$,
\begin{align}
    \mu_{1} - r_{0}^{2}  \frac{\partial \mu_{2}}{\partial t} + \nu \overline{\Delta^{*}} \mu_{2} - \frac{1}{a} \{ u_1, \mu_2 \} = 0.
\end{align}
The next calculation is shown below,
\begin{align*}
    D_{u_1}\mathcal{L}_{1} \bar{u} = & -\int_0^T \oint_{\partial \mcdomain}\int_{\Gamma} \lambda_4\rho \bar{u} dadbdydt + \int_{0}^T\int_C \int_{\Gamma}  \overline{{\Delta}^{*}} \mu_1 \bar{u}\rho da \bar{u} dbdy dt \\
    &+ \int_0^T \int_{\mcdomain}\int_{\Gamma} \left( \frac{1}{a}\{ \mu_{2}, u_{2} \} + \frac{2}{a^2} u_{2} \frac{\partial \mu_{2}}{\partial b} \right) \rho \bar{u} da db dy dt \\
    &+ \int_0^T \int_{\mcdomain}\int_{\Gamma} \left( \frac{1}{a}\{ \mu_{3}, u_{3} \} + \frac{2}{a^2} \left( u_{3} + r_{0} B_{0} (t) \right) \frac{\partial \mu_{3}}{\partial b} \right) \rho \bar{u} da db dy dt \\
    &+ \int_0^T \int_{\mcdomain}\int_{\Gamma} \frac{1}{a}\{ \mu_{4}, u_{4} \}\rho \bar{u} da db dy dt \\
    &- \int_0^T \int_{\mcdomain}\int_{\Gamma} \frac{1}{a}\{ u_{5}, \mu_{6} \}\rho \bar{u} da db dy dt \\
    &- \int_{\mcdomain} \int_{\Gamma} \frac{\alpha_1}{a^2} \left( \frac{\partial u_1}{\partial b} + a r_0 [v_d]_r \right) \frac{\partial \bar{u}}{\partial b} \rho da db dy \\
    &- \int_{\mcdomain} \int_{\Gamma} \frac{\alpha_1}{a^2} \left( \frac{\partial u_1}{\partial a} - a r_0 [v_d]_z \right) \frac{\partial \bar{u}}{\partial a} \rho da db dy . 
\end{align*}
Now it holds that $D_{u_1}\mathcal{L}_{1} \bar{u}=0$ if and only if for all $w \in H^1([0,T],H^1(\mcdomain)) \otimes L_{\rho}^{2} (\Gamma)$,
\begin{align*}
  0 &= \mathbb{E} \left[\lambda_4 \right]  \quad &\text{on } \partial \mcdomain \times [0,T] \\
  0 &= \mathbb{E} \left[ \overline{\Delta}^{*} \mu_{1} \right] + \mathbb{E} \left[\frac{1}{a} \{ \mu_{2}, u_2 \} + \frac{2}{a^2} u_2 \frac{\partial \mu_{2}}{\partial b} \right] \\
  & \quad + \mathbb{E} \left[\frac{1}{a} \{ \mu_{3}, u_3 \} + \frac{2}{a^2} (u_3 + r_{0} B_{0} (t)) \frac{\partial \mu_{3}}{\partial b} \right] \nonumber\\
  & \quad + \mathbb{E} \left[\frac{1}{a} \{ \mu_{4}, u_4 \} \right] - \mathbb{E} \left[\frac{1}{a} \{ u_{5}, \mu_{6} \} \right]  \quad & \text{on } \mcdomain \times [0,T] \\
  0 &= \mathbb{E} \left[\frac{\alpha_1}{a^2} \left( \frac{\partial u_1}{\partial b} + a r_0 [v_d]_r \right)+ \frac{\alpha_1}{a^2} \left( \frac{\partial u_1}{\partial a} - a r_0 [v_d]_z \right) \right] \quad & \text{on } \mcdomain
\end{align*}
The rest of the calculations are done in a similar manner and skipped for brevity.
\appendix
\section{Calculations for Transformations into Cylindrical Coordinates}\label{sec:appcalc}

\subsection{Derivations of EM fields}
In this part of the Appendix we derive the equations describing the substitutions \eqref{eq:alfven}-\eqref{eq:alfvenomega}.

Construction of $\psi$:
Recall that the PDE system includes~\eqref{eq:m1}, or $\nabla \cdot \mvec{v} = 0$ . In the cylindrical coordinates system, this condition becomes,
\begin{eqnarray*}
\frac{1}{r} \frac{\partial v_r}{\partial r} + \frac{1}{r} \frac{\partial v_{\phi}}{\partial \phi} + \frac{\partial v_z}{\partial z} = 0. \nonumber
\end{eqnarray*}
But, due to $\phi$ invariance, $\frac{\partial v_{\phi}}{\partial \phi} = 0$. So, the above equation becomes,
\begin{equation} 
\frac{1}{r} \frac{\partial r v_r}{\partial r} + \frac{\partial v_z}{\partial z} = 0. \nonumber
\end{equation}
 Recalling the form of $\psi$ and writing the surface integration $dS = r ~ dr ~ d\phi ~ $, we get,
\begin{equation} \label{psir}
\psi (r) = \frac{1}{2 \pi} \int_{0}^{r} \int_{0}^{2 \pi} v_{z} (\rho, z) \rho d\rho ~ d\phi = \int_{0}^{r} v_{z} (\rho, z) \rho ~ d\rho. 
\end{equation}
The above expression holds as $v_{z} (r,z)$ is independent of $\phi$, which is true because of $\phi$ invariant assumption. Now, we get,
\begin{eqnarray}\label{vz}
\frac{\partial }{\partial r} \psi = r v_{z} (r,z). \Longrightarrow v_{z} = \frac{1}{r} \frac{\partial \psi}{\partial r}.
\end{eqnarray}

Now, differentiating \eqref{psir} with respect to $z$ we obtain,
\begin{eqnarray}
    \frac{\partial \psi}{\partial z} = \int_{0}^{r} \frac{\partial v_{z}}{\partial z} \rho ~ d \rho. \nonumber
\end{eqnarray}
After substituting the expression of \eqref{vz} in the above equation, we get,
\begin{eqnarray}
    \frac{\partial \psi}{\partial z} = - \int_{0}^{r} \frac{1}{\rho} \frac{\partial \rho v_{\rho}}{\partial \rho} \rho ~ d \rho = - r v_{r}. \nonumber
\end{eqnarray}
from which we obtain the expression
\begin{eqnarray}
    v_{r} = - \frac{1}{r} \frac{\partial \psi}{\partial z}. \nonumber
\end{eqnarray}
We observe that $\nabla \phi = \frac{1}{r} \hat{i}_{\phi}$. The poloidal part of the velocity component becomes,
\begin{eqnarray}
    v_{p} = v_{r} \hat{i}_{r} + v_{z} \hat{i}_{z} = \nabla \psi \times \nabla \phi . \nonumber
\end{eqnarray}
Now, let us compute the expression of the toroidal  directional component of the velocity $v_{\phi}$. With the help of the property $\nabla \cdot \nabla \times \mvec{v} = 0$ we can write the vorticity $\omega = \nabla \times \mvec{v}$ as follows,
\begin{eqnarray} \label{omega1}
    \omega_{r} =  - \frac{1}{r} \frac{\partial f}{\partial z}, ~ \omega_{z} =  \frac{1}{r} \frac{\partial f}{\partial r}, \nonumber
\end{eqnarray}
for some scalar function $f$. Now, from the condition of $\nabla \times \mvec{v} = \omega$, we get,
\begin{eqnarray}
&&
    - \frac{\partial v_{\phi}}{\partial z} = \omega_{r}, \\
    && \label{omega2}
 \frac{1}{r}   \frac{\partial r v_{\phi}}{\partial r} = \omega_{z}, \\
 &&
 \frac{\partial v_{r}}{\partial z} - \frac{\partial v_{z}}{\partial r} = \omega_{\phi}. \nonumber
\end{eqnarray}
Equations \eqref{omega1} and \eqref{omega2} together imply
\begin{eqnarray}
    \frac{1}{r}   \frac{\partial r v_{\phi}}{\partial r} = \frac{1}{r}   \frac{\partial f}{\partial r}. \nonumber
\end{eqnarray}
suggesting $v_{\phi} = \frac{f}{r}$ to be a solution. The velocity field has the following expression,
\begin{eqnarray} \label{eq:useful1}
    \mvec{v}(r,z) = \nabla \psi \times \nabla \phi + v_{\phi} \hat{i}_{\phi}. 
\end{eqnarray}
Similarly, we get the following expression for the magnetic field as,
\begin{eqnarray} \label{eq:useful2}
    \mvec{B}(r,z) = \nabla \chi \times \nabla \phi + B_{\phi} \hat{i}_{\phi}, 
\end{eqnarray}
for some scalar function $\chi$. 

\subsection{Derivations for current and vorticity}
Recall the following expressions for the plasma current and vorticity,
\begin{eqnarray} &&
    \nabla \times \mvec{B} = \mvec{J_{p}}, \nonumber \\
    &&
    \nabla \times \mvec{v} = \mvec{\omega}. \nonumber
 \end{eqnarray}
 Incorporating \eqref{eq:useful1} and \eqref{eq:useful2} into these expressions yields
 \begin{eqnarray}
&&
\mvec{\omega} =  \nabla \times (\nabla \psi \times \nabla \phi + v_{\phi} \hat{i}_{\phi}), \nonumber \\
&&
\mvec{J_{p}} = \nabla \times (\nabla \chi \times \nabla \phi + B_{\phi} \hat{i}_{\phi}). \nonumber
 \end{eqnarray}
 Rearranging terms,
 \begin{eqnarray}
\nabla \times (\nabla \chi \times \nabla \phi + B_{\phi} \hat{i}_{\phi}) & = & \nabla \times (B_{\phi} \hat{i}_{\phi}) + \nabla \times (\nabla \chi \times \nabla \phi). \nonumber
 \end{eqnarray}
Now, $\nabla \times (B_{\phi} \hat{i}_{\phi}) = - \frac{1}{r} \frac{\partial (r B_{\phi})}{\partial z} \hat{i}_{r} + \frac{1}{r} \frac{\partial (r B_{\phi})}{\partial r} \hat{i}_{z} = \nabla (r B_{\phi}) \times \nabla \phi$. For the second term,
\begin{eqnarray}
    \nabla \times (\nabla \chi \times \nabla \phi) = \nabla \times \left(\nabla \chi \times \frac{\hat{i_{\phi}}}{r}\right). \nonumber
\end{eqnarray}
Now, \begin{align}
\nabla \chi \times \frac{\hat{i_{\phi}}}{r} &= - \frac{1}{r} \frac{\partial \chi}{\partial z} \hat{i}_{r} + \frac{1}{r} \frac{\partial \chi}{\partial r} \hat{i}_{z} .
\label{eq:identity}
\end{align} 
which implies
\begin{eqnarray}
    \nabla \times \left(- \frac{1}{r} \frac{\partial \chi}{\partial z} \hat{i}_{r} + \frac{1}{r} \frac{\partial \chi}{\partial r} \hat{i}_{z} \right) & = & 0 \cdot \hat{i}_{r} + 0 \cdot \hat{i}_{z} \nonumber \\
    &&
    + \Big[- \frac{1}{r} \frac{\partial^{2} \chi}{\partial z^{2}} - \frac{\partial \chi}{\partial r} \Big (\frac{1}{r} \frac{\partial \chi}{\partial r} \Big) \Big] \hat{i}_{\phi} \nonumber \\
    & = &
    \Big[- \frac{1}{r} \frac{\partial^{2} \chi}{\partial z^{2}} + \frac{1}{r^{2}} \frac{\partial \chi }{\partial r} - \frac{1}{r} \frac{\partial^{2} \chi}{\partial r^{2}} \Big] \hat{i}_{\phi}, \nonumber \\
    & = &
   - \Delta^{*} \chi \nabla \phi, \nonumber
\end{eqnarray}
where, $\Delta^{*} f = \frac{\partial^{2} f}{\partial z^{2}} + \frac{\partial^{2} f}{\partial r^{2}} - \frac{1}{r} \frac{\partial f}{\partial r}$. Therefore, we get,
\begin{eqnarray}
    \mvec{J_{p}} = \Big[ \nabla (r B_{\phi}) \times \nabla \phi - {\Delta}^{*} \chi \nabla \phi \Big]. \nonumber
\end{eqnarray}
Pursuing the same line of argument for $\omega$ yields the analogous expression
\begin{eqnarray}
    \mvec{\omega} = \Big[ \nabla (r v_{\phi}) \times \nabla \phi - {\Delta}^{*} \psi \nabla \phi \Big]. \nonumber
\end{eqnarray}

\subsection{Derivations for cylindrical boundary conditions}\label{sec:appcylbound}
Now, \eqref{eq:rectbound} are transformed into cylindrical coordinates to get the boundary conditions in cylindrical coordinates. These sets of equations do not appear in the main text, as they are an intermediate step between the rectangular and final coordinate bases.  
\begin{eqnarray} &&
    v|_{t = 0} = \nabla \psi_{0} \times \nabla \phi + v_{0} \hat{i}_{\phi} \to  \nabla \psi|_{t = 0} \times \nabla \phi + {v_{\phi}}(0) \hat{i}_{\phi} = \nabla \psi_{0} \times \nabla \phi + v_{0} \hat{i}_{\phi}. \nonumber
\end{eqnarray} Take, $v_{\phi} = {v_{0}}$ and $\psi|_{t = 0} = \psi_{0}$. Similarly, $B|_{t = 0} = \nabla \chi_{0} \times \nabla \phi + B_{1} \hat{i}_{\phi}$ yields, $B_{\phi}(0) + B_{0} (0) \frac{r_{0}}{r} = {B_{1}}$ and $\chi|_{t = 0} = \chi_{0}$. The condition, $v = 0$ yields $\psi = 0, v_{\phi} = 0$. For the last two conditions, we note that $\partial\Omega$ is spanned by $\hat{i}_{\phi}, \hat{i}_z$ and $n$ is spanned by $\hat{i}_{r}$. Thus, $B \cdot n = B_{r} = - \frac{1}{r} \frac{\partial \chi}{\partial z}$ and $B \cdot n = 0 \to  \frac{\partial \chi}{\partial z} = 0$. Now, $(\nabla \times B)|_{\partial\Omega} = 0 \to  {J_{p}}|_{\partial\Omega} = 0 \to  {J_{p}}_{\phi} = 0 ~ \& ~ {J_{p}}_{z} = 0$. So, we get, $\frac{\partial r B_{\phi}}{\partial r} = 0$ and ${\Delta}^{*} \chi = 0$. Thus, we get, $B_{\phi} + r \frac{\partial B_{\phi}}{\partial r} = 0$ and $\frac{\partial^{2} \chi}{\partial r^{2}} - \frac{1}{r} \frac{\partial \chi}{\partial r} = 0$.

\subsection{Derivations of the plasma equations transformed by~\eqref{eq:alfven2}} \label{sec:apptran} 
After equating the toroidal components, i.e. the $\hat{i}_{\phi}$ directional parts of the  \eqref{eq:alfvenomega} and applying \eqref{eq:alfven}, we get,
\begin{align}
    {\Delta}^{*} \psi &= - r \omega_{\phi}.
    \label{eq:reduced1}
\end{align}
We have the following coordinate transformations, \begin{align}
    a = \frac{r}{r_{0}}, ~ b = \frac{z}{r_{0}}, ~ u_{1} = \frac{\psi}{r_{0}}, ~ u_{2} = r_{0} r \omega_{\phi}, ~ \overline{\Delta}^{*} f = \frac{\partial^{2} f}{\partial a^{2}} - \frac{1}{a} \frac{\partial f}{\partial a} + \frac{\partial^{2} f}{\partial b^{2}}, \nonumber
\end{align}
The Laplacian operator becomes
\begin{align}
    \overline{\Delta}^{*} &= r_{0}^{2} {\Delta}^{*}.  \nonumber
\end{align}
Thus, after multiplying \eqref{eq:reduced1} by $r_{0}$, we finally get,
\begin{align}
    \overline{{\Delta}}^{*} \frac{\psi}{r_{0}} &= - r r_{0} \omega_{\phi} \to  \overline{{\Delta}}^{*} u_{1} = - u_{2}.  \nonumber
\end{align}
Considering the $\phi$ component of~\eqref{eq:alfvencurrent}, we obtain
\begin{align}
    {\Delta}^{*} \chi &= - r {J_{p}}_{\phi},
    \label{eq:current} 
\end{align}
where ${J_{p}}_{\phi}$ is obtained from \eqref{eq:lorentz}, substituting~\eqref{eq:bcylin} and~\eqref{eq:ecylinphi} to obtain
\begin{align}
    \eta {J_{p}}_{\phi}  &= \frac{1}{r} \left( \frac{\partial \chi}{\partial t} + \kappa^{'} (t) \right) + (\nabla \psi \times \nabla \phi) \times (\nabla \chi \times \nabla \phi), \nonumber \\
                  &= \frac{1}{r} \left( \frac{\partial \chi}{\partial t} + \kappa^{'} (t) \right) - \frac{\nabla \chi \times \nabla \psi}{r} \cdot \nabla \phi.
                  \label{eq:eta}  
\end{align}
With the following coordinate transformations,
\begin{align}
    u_{6} &= r_{0} r {J_{p}}_{\phi}, ~ u_{5} = \frac{\chi}{r_{0}},  \nonumber 
\end{align}
we get the following transformed version of the equation \eqref{eq:current} with $\Delta^{*} \bar{\chi} = \Delta^{*} \chi$,
\begin{align}
    \overline{{\Delta}^{*}} u_{5} &= - u_{6} .  \nonumber
\end{align}
From expression \eqref{eq:eta} we get,
\begin{align}
    \eta u_{6}  &= r_{0}^{2} \frac{\partial u_{5}}{\partial t} + r_{0} \kappa^{'} (t) + \frac{1}{a} \{ u_{5}, u_{1} \},  \nonumber
\end{align}
where recall that the Poisson bracket is denoted $\{ f,g \} = \frac{\partial f}{\partial a} \frac{\partial g}{\partial b} - \frac{\partial f}{\partial b} \frac{\partial g}{\partial a} $.
Taking curl of \eqref{eq:max2} and applying the angular symmetry, we get,
\begin{align}
    \nu \nabla^{2} \omega &= \frac{\partial \omega}{\partial t} + \nabla \times \left(\omega \times v - J_{p} \times (B+B_{\kappa}^{ind}) \right).
    \label{eq:vorticity}
\end{align}
Now, taking the toroidal component of \eqref{eq:vorticity}, we get,
\begin{align}
 \Big( \frac{\partial \omega}{\partial t} \Big)_{\phi}    &= \frac{\partial (r \omega_{\phi})}{\partial t} \nabla \phi , \nonumber \\
 \Big( \nabla^{2} \omega \Big)_{\phi} &= \Big( \frac{1}{r} \frac{\partial}{\partial r} \Big[ r \frac{\partial \omega_{\phi}}{\partial r}\Big] + \frac{\partial^{2} \omega_{\phi}}{\partial z^{2}} - \frac{\omega_{\phi}}{r^{2}} \Big) \hat{i}_{\phi}, \nonumber \\
 \Delta^{*} (r \omega_{\phi}) &= \frac{\partial^{2} (r \omega_{\phi})}{\partial r^{2}} - \frac{1}{r} \frac{\partial (r \omega_{\phi})}{\partial r} + \frac{\partial^{2} (r \omega_{\phi})}{\partial z^{2}} , \nonumber \\
 \to  \Delta^{*} (r \omega_{\phi})  \nabla \phi &= \Big( \nabla^{2} \omega \Big)_{\phi}, \nonumber \\
 \Big( \nabla \times (\omega \times v) \Big)_{\phi} &= \nabla \times \Big[ (\nabla (r v_{\phi}) \times \nabla \phi + \omega_{\phi} \hat{i}_{\phi}) \times (\nabla \psi \times \nabla \phi + v_{\phi} \hat{i}_{\phi}) \Big]_{\phi}, \nonumber \\
 &= \nabla \times \Big[ (\nabla (r v_{\phi}) \times \nabla \phi) \times  v_{\phi} \hat{i}_{\phi} + \omega_{\phi} \hat{i}_{\phi} \times (\nabla \psi \times \nabla \phi) \Big]_{\phi}.
\end{align}
We use the following derivations to get the final equality,
\begin{align}
\nabla \times \left((\nabla (r v_{\phi}) \times \nabla \phi) \times  v_{\phi} \hat{i}_{\phi} \right) &= - \nabla \times \Big( \frac{v_{\phi}}{r} \Big [\frac{\partial (r v_{\phi})}{\partial r} \hat{i}_{r} + \frac{\partial (r v_{\phi})}{\partial z} \hat{i}_{z} \Big]
 \Big), \nonumber \\
 &= \left(- \frac{\partial}{\partial z} \left[  \frac{v_{\phi}}{r} \frac{\partial (r v_{\phi})}{\partial r} \right] +  \frac{\partial}{\partial r} \left[  \frac{v_{\phi}}{r} \frac{\partial (r v_{\phi})}{\partial z}  \right] \right) \hat{i}_{\phi}, \nonumber \\
 &= \nabla (r v_{\phi}) \times \nabla \left(\frac{v_{\phi}}{r} \right).
 \label{eq:useful}
\end{align}
With \eqref{eq:useful} we get the following expression,
\begin{align}
    \Big( \nabla \times (\omega \times v) \Big)_{\phi} &= \nabla (r v_{\phi}) \times \nabla \left(\frac{v_{\phi}}{r}\right) + \nabla \left(\frac{\omega_{\phi}}{r}\right) \times \nabla \psi.
    \label{eq:curlidentity1}
\end{align}
And similarly, we get,
\begin{align}
    \Big( \nabla \times \left( J_{p}  \times (B + B_{\kappa}^{int}) \right) \Big)_{\phi} &= \nabla (r B_{\phi}) \times \nabla \left(\frac{B_{\phi}}{r} + \frac{B_{0}r_{0}}{r^{2}} \right) \nonumber \\
    & \quad + \nabla \left(\frac{{J_{p}}_{\phi}}{r} \right) \times \nabla \chi.
    \label{eq:curlidentity2} 
\end{align}
After arranging the equation \eqref{eq:max2}, with the following identity,
\begin{align}
    \frac{1}{2} \nabla v \cdot v &= (v \cdot \nabla) v + v \times \omega, \nonumber
\end{align}
and taking the toroidal component of \eqref{eq:max2}, we get,
\begin{align}
   \nu \Big( \nabla^{2} v \Big)_{\phi} &= \Big( \frac{\partial v}{\partial t} \Big)_{\phi} + \nabla \Big( p + \frac{1}{2} v \cdot v  \Big)_{\phi} + \Big ( \omega \times v - J_{p} \times (B + B_{\kappa}^{ind}) \Big)_{\phi}.
   \label{eq:transform2}
\end{align}
Performing the appropriate coordinate transformations,
\begin{align}
    \Big( \frac{\partial v}{\partial t} \Big)_{\phi}    &= \frac{\partial (r v_{\phi})}{\partial t} \nabla \phi , 
    \label{eq:ohm1} \nonumber \\
 \Big( \nabla^{2} v \Big)_{\phi} &= \Big( \frac{1}{r} \frac{\partial}{\partial r} \Big[ r \frac{\partial v_{\phi}}{\partial r}\Big] + \frac{\partial^{2} v_{\phi}}{\partial z^{2}} - \frac{v_{\phi}}{r^{2}} \Big) \hat{i}_{\phi}, \nonumber \\
 \Delta^{*} (r v_{\phi}) &= \frac{\partial^{2} (r v_{\phi})}{\partial r^{2}} - \frac{1}{r} \frac{\partial (r v_{\phi})}{\partial r} + \frac{\partial^{2} (r v_{\phi})}{\partial z^{2}} , \nonumber \\
 \to  \Delta^{*} (r v_{\phi}) \nabla \phi &= \Big( \nabla^{2} v \Big)_{\phi},  \\
 \nabla \Big( p + \frac{1}{2} v \cdot v  \Big)_{\phi} &= 0, \nonumber \\
 \Big( \omega \times v  \Big)_{\phi} &= \frac{1}{r^{2}} \Big[ \nabla (r v_{\phi}) \times \nabla \psi \Big] \hat{i}_{\phi}, ~ ( see \eqref{eq:identity}) \nonumber \\
 \Big( J_{p} \times (B + B_{\kappa}^{ind}) \Big)_{\phi} &= \frac{1}{r^{2}} \Big[ \nabla (r B_{\phi}) \times \nabla \chi \Big] \hat{i}_{\phi}. ~ (see \eqref{eq:identity}) \nonumber
\end{align}
Finally, we take the curl on both sides of the equation \eqref{eq:max3} and then take its toroidal part along with the identities \eqref{eq:ohm1}, \eqref{eq:curlidentity1}, \eqref{eq:curlidentity2} and $\nabla \times (\nabla f \times \nabla \phi) = - \Delta^{*} f \nabla \phi$,
\begin{align}
    - \frac{\partial (r B_{\phi})}{\partial t} \nabla \phi - \Big[ \nabla (\chi) \times \nabla \Big( \frac{v_{\phi}}{r} \Big) + \nabla \Big( \frac{B_{\phi}}{r} + \frac{B_{0} r_{0}}{r^{2}} \Big) \times \nabla \psi  \Big]  &= - \eta \Delta^{*} (r B_{\phi}) \nabla \phi . 
    \label{eq:transform3}
\end{align}
We use the following coordinate transformations,
\begin{align}
    u_{3} = r B_{\phi}, ~ u_{4} = r v_{\phi} \nonumber
\end{align}
to transform the equations \eqref{eq:transform2}, \eqref{eq:transform3} and \eqref{eq:vorticity}, respectively, as follows,
\begin{align}
    \nu \overline{\Delta^{*}} u_{4} &= r_{0}^{2} \frac{\partial u_{4}}{\partial t} + \frac{1}{a} \Big( \{ u_{3}, u_{5} \} + \{ u_{1}, u_{4} \} \Big) \nonumber \\
    \eta \overline{\Delta^{*}} u_{3} &= r_{0}^{2} \frac{\partial u_{3}}{\partial t} + r_{0}^{3} B_{0}^{'}(t) + \frac{1}{a} \Big( \{ u_{4}, u_{5} \} + \{ u_{1}, u_{3} \} \Big) \nonumber \\
    & \quad + \frac{2}{a^{2}} \Big( \left(u_{3} + r_{0} B_{0} (t) \right) \frac{\partial u_{1}}{\partial b}  - u_{4} \frac{\partial u_{5}}{\partial b} \Big),   \nonumber \\
    \nu \overline{\Delta^{*}} u_{2} &= r_{0}^{2} \frac{\partial u_{2}}{\partial t} + \frac{1}{a} \{ u_{1}, u_{2} \} + \frac{2}{a^{2}} u_{2} \frac{\partial u_{1}}{\partial b} + \frac{1}{a} \{ u_{6} , u_{5} \} - \frac{2}{a^{2}} u_{6}  \frac{\partial u_{5}}{\partial b} \nonumber \\
    & \quad + \frac{2}{a^{2}} \Big[ (u_{3} + r_{0} B_{0} (t)) \frac{\partial u_{3}}{\partial b} - u_{4} \frac{\partial u_{4}}{\partial b} \Big].  \nonumber
\end{align}

\subsection{Boundary Conditions}\label{sec:apptranbound}
Following Section~\ref{sec:appcylbound}, the boundary conditions in the cylindrical coordinates system are described as,
\begin{eqnarray}
&& \label{eq:boundaryrb1}
\psi|_{t = 0} = \psi_{0}, ~ v_{\phi}|_{t = 0} = v_{0}, ~ \chi|_{t = 0} = \chi_{0}, ~ B_{\phi}|_{t = 0} + B_{0} (0) \frac{r_{0}}{r} = B_{1} \text{~ on $\mcdomain$}, \\
&& \label{eq:boundaryrb2}
\psi|_{\partial \mcdomain} = 0, ~ {v_{\phi}}|_{\partial \mcdomain} = 0 \text{~ in $\partial \mcdomain \times [0,T]$}, \\
&& \label{eq:boundaryrb3}
B_{\phi} + r \frac{\partial B_{\phi}}{\partial r} = 0, ~ \frac{\partial^{2 } \chi}{\partial r^{2}} - \frac{1}{r} \frac{\partial \chi}{\partial r} = 0 \text{~ on $\partial \mcdomain \times [0,T]$}, \\
&& \label{eq:boundaryrb4}
\frac{\partial \chi}{\partial z} = 0 \text{~ on $\partial \mcdomain \times [0,T]$}.
\end{eqnarray}
Under the assumption that initial plasma velocity is zero, the reduced form of the boundary conditions \eqref{eq:boundaryb1}-\eqref{eq:boundaryb4} are described as,
\begin{align}
    \frac{\partial u_{5}}{\partial b} &= 0, & \text{On }\partial \mcdomain \times [0,T] \nonumber \\
    \frac{\partial u_{3}}{\partial a} = 0, ~ \Big[\frac{\partial^{2} u_{5}}{\partial a^{2}} - \frac{1}{a} \frac{\partial u_{5}}{\partial a} \Big] &= 0, & \text{On }\partial \mcdomain \times [0,T] \nonumber \\
    u_{1}, ~ u_{4}  &= 0, & \text{On }\partial \mcdomain \times [0,T] \nonumber \\
  \left(\psi_{0} = 0, v_{0} = 0\right) \to   \mvec{u}_{1}(0), ~ \mvec{u}_{4}(0) &= 0, & \text{On }\mcdomain  \nonumber \\
  u_{5} = A:=\frac{\chi_{0}}{r_{0}}, ~ u_{3} &= r B_{1} - r_{0} B_{0} (0). ~
   & \text{On }\mcdomain  \nonumber
\end{align}

\nocite{*}
\bibliographystyle{unsrt}
\bibliography{refs}
\bigskip
\end{document}